\documentclass[aps,prl,twocolumn,preprintnumbers,amsmath,amssymb,superscriptaddress,10pt]{revtex4-2}
\usepackage{amsmath,graphicx}
\usepackage[utf8]{inputenc}
\usepackage[T1]{fontenc}
\usepackage{xcolor}
\usepackage{textcomp}
\usepackage{bm}

\usepackage{physics}
\usepackage{amsmath}
\usepackage{tikz}
\usepackage{mathdots}
\usepackage{yhmath}
\usepackage{cancel}
\usepackage{color}
\usepackage{array}
\usepackage{multirow}
\usepackage{amssymb}
\usepackage{gensymb}
\usepackage{tabularx}
\usepackage{extarrows}
\usepackage{booktabs}
\usetikzlibrary{fadings}
\usetikzlibrary{patterns}
\usetikzlibrary{shadows.blur}
\usetikzlibrary{shapes}

\usepackage{mathtools}

\usepackage{color}
\definecolor{LinkColor}{rgb}{0.75,0.0,0.2}

\usepackage{hyperref}
\hypersetup{
	pdfauthor={good guys},
	pdftitle={good title},
	colorlinks=true,
	citecolor=LinkColor,
	linkcolor=LinkColor,
	urlcolor=LinkColor,
}

\usepackage{listings}
\definecolor{lightgray}{gray}{1}

\usepackage{theorem}

\usepackage{tikz}

\newcommand{\nc}{\newcommand}
\nc{\braoprket}[3]{\langle#1|#2|#3\rangle}
\nc{\opn}[1]{\operatorname{#1}}
\nc{\avg}[1]{\langle#1\rangle}
\nc{\ketbrasame}[1]{|#1\rangle\!\langle#1|}
\nc{\swap}{\opn{SWAP}}
\nc{\E}{\mathbb{E}}
\nc{\Var}{\opn{Var}}
\nc{\dg}{\dagger}

\newcommand{\panel}[3]{%
  \begin{tikzpicture}[inner sep=0pt, outer sep=0pt]
    \node[anchor=south west, inner sep=1.5pt] (img) at (0,0)
      {\includegraphics[width=\dimexpr #2-3pt\relax]{#1}}; 
    \node[anchor=north west, font=\bfseries, fill=white, rounded corners=0pt, inner sep=0pt]
      at ([xshift=0pt,yshift=-0.1pt]img.north west) {#3};
  \end{tikzpicture}%
}

\newcommand{\panelH}[3]{
	\begin{tikzpicture}[inner sep=0pt, outer sep=0pt,
	baseline=(current bounding box.north)]
	\node[anchor=south west, inner sep=1.5pt] (img) at (0,0)
	{\includegraphics[height=\dimexpr #2-3pt\relax,keepaspectratio]{#1}};
	\path ([yshift=0.6ex]img.north west) -- ([yshift=0.6ex]img.north east);
	\node[anchor=north west, font=\bfseries, fill=white,
	inner sep=0.6pt, text height=1.3ex, text depth=0.25ex]
	at ([yshift=-0.4ex]img.north west) {#3};
	\end{tikzpicture}%
}

\newcommand{\twopanel}[4]{%
  \begin{tikzpicture}[inner sep=0pt, outer sep=0pt]
    \node[anchor=south west, inner sep=1.5pt] (img) at (0,0)
      {\includegraphics[width=\dimexpr #2-3pt\relax]{#1}}; 

    \node[anchor=north west, font=\bfseries, fill=white, rounded corners=0pt, inner sep=0pt]
      at ([xshift=0pt,yshift=-0.1pt]img.north west) {#3};

    \node[anchor=north west, font=\bfseries, fill=white, rounded corners=0pt, inner sep=0pt]
      at ([xshift=0.5\linewidth,yshift=-0.1pt]img.north west) {#4};
  \end{tikzpicture}%
}

\usepackage[normalem]{ulem}

\begin{document}

\title{Topologically Enforced Lifshitz Multicriticality in One Dimension}

\author{Kuang-Hung Chou}
\affiliation{Department of Physics, National Tsing Hua University, Hsinchu 30013, Taiwan}

\author{Xue-Jia Yu}
\email{xuejiayu@eitech.edu.cn}
\affiliation{Eastern Institute of Technology, Ningbo 315200, China}

\begin{abstract}
Recent advances have revealed that topology can further enrich the universality classes of quantum phase transitions, thereby extending beyond the traditional paradigms of statistical and condensed matter physics. However, multicriticality between topologically distinct quantum critical lines remains insufficiently explored. In this Letter, we systematically construct and investigate a novel class of topologically enforced Lifshitz multicritical points in one-dimensional chiral-symmetric fermionic systems. Such multicriticality is driven solely by changes in the topology of neighboring critical lines, beyond previously recognized multicritical points that are typically induced by changes in critical exponents. More importantly, the topologically enforced multicriticality identified here can host robust topological degeneracies while surprisingly exhibiting a breakdown of the Li–Haldane bulk-boundary correspondence—a phenomenon we elucidate through a simple physical picture. 

\end{abstract}

\maketitle

\emph{Introduction.}---Phase transitions and their associated universality classes are among the most fundamental concepts in modern physics. Traditionally, the universality class of a phase transition has long been believed to be fully determined by a set of critical exponents, forming the standard paradigm for understanding phase transitions over the past decades~\cite{landau2013statistical,Sachdev_1999,Sondhi1997RMP}. However, this conventional picture has recently been challenged by the discovery of topological physics in  quantum phase transitions~\cite{YU20261,Scaffidi2017PRX,Verresen2018PRL,Parker2018PRB,verresen2020topologyedgestatessurvive,Verresen2021PRX,Duque2021PRB_L,Thorngren2021PRB,Yu2022PRL,Hidaka2022PRB,Ye2022SciPost,Mondal2023PRB,Wen2023PRB,Yu2024PRL,Li2024SciPost,Yu2024PRB,Su2024PRB,Zhong2024PRA,Zhang2024PRA,Choi2024PRB,Saran2024PRB_L,Li2025SciPost,Yang2025CP,Zhong2025PRB,Wen2025PRB,Li2025PRB_L,Rey2025PRB,Zhou2025CP,Cardoso2025PRB,Flores2025PRL,Jia2025PRL,Huang2025SciPost,kumar2025topologicallynontrivialmulticriticalpoints,chou2025ptsymmetryenrichednonunitarycriticality,guo2025generalizedlihaldanecorrespondencecritical,banerjee2025entanglementspectrumgaplesstopological,prembabu2025multicriticalitypurelygaplessspt,prembabu2025noninvertibleinterfacessymmetryenrichedcritical,deng2025anomalousdynamicalscalingtopological,tan2025exploringnontrivialtopologyquantum,yang2026topologicalquantumcriticalityquasiperiodic,Yang2026PRB_L}, in which topology can further enrich the corresponding universality classes even when they share the same critical exponents. As a consequence, topologically nontrivial quantum criticality emerges beyond the traditional Landau-Ginzburg symmetry-breaking paradigm. More importantly, these conceptual advances are not trivial generalizations of topology to quantum critical points, but instead give rise to fundamentally new phenomena, including unconventional topological invariants~\cite{Yu2022PRL,Parker2018PRB,Zhou2025CP}, generalized bulk-boundary correspondence~\cite{Yu2024PRL,Zhong2025PRB,guo2025generalizedlihaldanecorrespondencecritical}, algebraically localized edge modes~\cite{Verresen2021PRX,Yang2025CP}, anomalous dynamical scaling behaviors~\cite{deng2025anomalousdynamicalscalingtopological}, and intrinsically gapless topological states~\cite{Thorngren2021PRB,Zhang2024PRA}.

On a different front, multicritical points (MCPs) are special points that require tuning at least two parameters, as they lie at the intersection of multiple phase transition lines. They serve as key organizing centers for complex phases of matter, ranging from the tricritical point in the classical liquid–gas phase diagram to a broad range of modern physics research topics, including lattice gauge theory~\cite{Tupitsyn2010PRB,Somoza2021PRX,Bonati2022PRB}, competing order parameters in magnetism and superconductivity~\cite{Classen2016PRB,Roy2018PRX,Herbut2022PRB}, superradiant transitions~\cite{Soriente2018PRL,Zhu2020PRL}, and deconfined criticality~\cite{Zhao2020PRL,Lu2021PRB,Chester2024PRL,Chen2024PRL}. The recent discovery of topological classification in quantum phase transitions provides a new route toward realizing MCPs~\cite{Kumar2021SR,Wang2022SciPost,Kumar2023PRB,Wang2023PRB_L,Yu2024PRB,Zhou2025CP}, whereas many previously recognized ones arise at the intersection of distinct phase transition lines characterized by different critical exponents. This naturally raises a fundamental and intriguing question: Can we systematically construct a class of MCPs induced solely by changes in the topology of neighboring critical lines, and more importantly, can such topologically enforced MCPs exhibit fundamentally new physics beyond previously recognized ones?

In this Letter, we fill this gap by systematically investigating topologically enforced Lifshitz MCPs in a class of one-dimensional chiral-symmetric fermionic models. Specifically, we first provide a general lattice construction for realizing such novel MCPs, which are driven solely by changes in the topology of neighboring critical lines, in contrast to conventional MCPs that are instead associated with changes in critical exponents. More importantly, we further demonstrate that these MCPs can be topologically nontrivial—they exhibit robust topological degeneracies in the bulk entanglement spectrum while unexpectedly displaying a breakdown of the Li–Haldane bulk-boundary correspondence—a phenomenon unique to the topological multicriticality. We also provide a simple physical picture to explain the origin of this breakdown.


\emph{Model and phase diagram.}---We consider topologically distinct quantum critical lines realized by linear combinations of two competing Hamiltonians, $H_{\alpha}+H_{\alpha+1}$, where $H_{\alpha}=-\sum_j\left(b_j^\dagger a_{j+\alpha}+\mathrm{h.c.}\right)$. Here, $a_j$ and $b_j$ denote annihilation operators on the two sublattices $A$ and $B$ within unit cell $j$, while $\alpha$ specifies the hopping range. The $\alpha$-chain Hamiltonian $H_{\alpha}$ preserves chiral symmetry and belongs to the AIII symmetry class~\cite{Altland1997PRB}; in momentum space, a generic two-band chiral Hamiltonian can be written as $\mathcal H(k)=h_x(k)\sigma_x+h_y(k)\sigma_y$, where $\sigma_{x,y}$ are Pauli matrices in sublattice space. Equivalently, the Hamiltonian is characterized by the off-diagonal Bloch element $v_k=h_x(k)+i h_y(k)$. The topology of chiral systems is commonly characterized by the winding number $W=(2\pi)^{-1}\int_{\rm BZ} dk\,[-i v_k^{-1}\partial_k v_k]$. The quantum critical line separating gapped topological phases with winding numbers $\alpha$ and $\alpha+1$ has recently been argued to remain topologically nontrivial~\cite{verresen2020topologyedgestatessurvive,guo2025generalizedlihaldanecorrespondencecritical} and to host $\mathrm{min}\{2|\alpha|,2|\alpha+1|\}$ exponentially localized edge modes even at criticality; in the following, we refer to it as the $\alpha$ critical line. However, the ordinary winding number is ill-defined in this case. Instead, their topology can be diagnosed through the auxiliary complex function associated with $H_{\alpha}+H_{\alpha+1}$~\cite{Verresen2018PRL,Jones2019JSP} (see End Matter for details):
The number of topological edge modes is determined by the number of zeros inside the unit circle. The system becomes critical when a zero lies on the unit circle, while the corresponding critical point possesses dynamical exponent $z=m$ when the critical zero on the unit circle is $m$-fold degenerate.

\begin{figure}[t]
  \centering
  \begin{minipage}[b]{1.0\linewidth}
    \panel{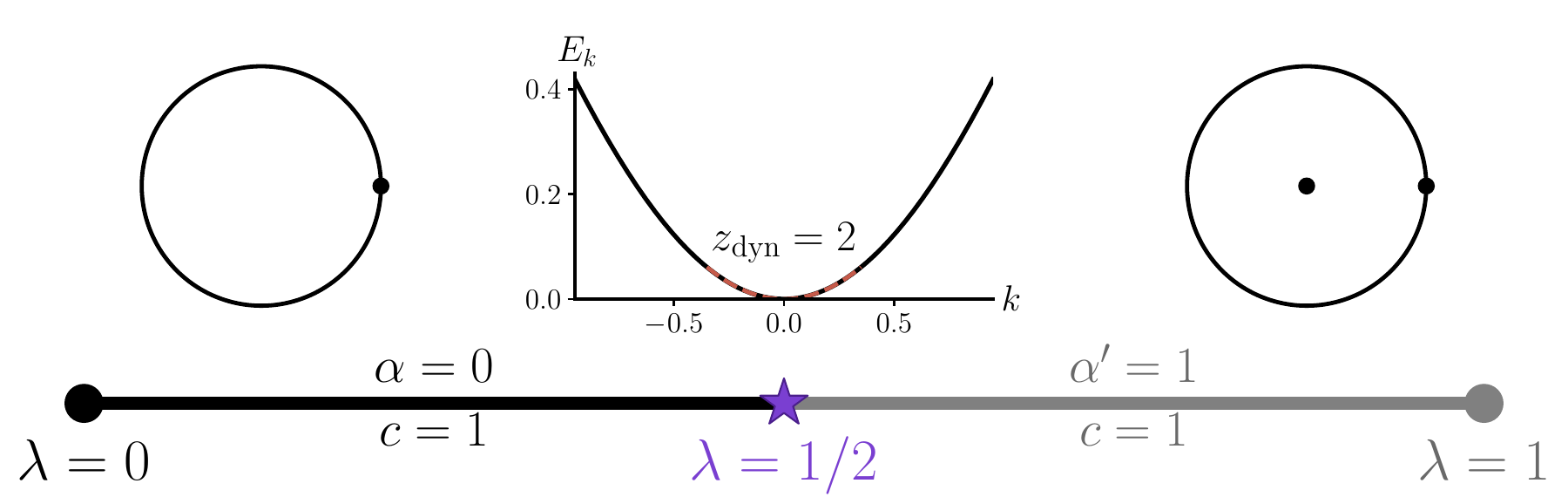}{\linewidth}{(a)}
  \end{minipage}
  \begin{minipage}[b]{1.0\linewidth}
    \panel{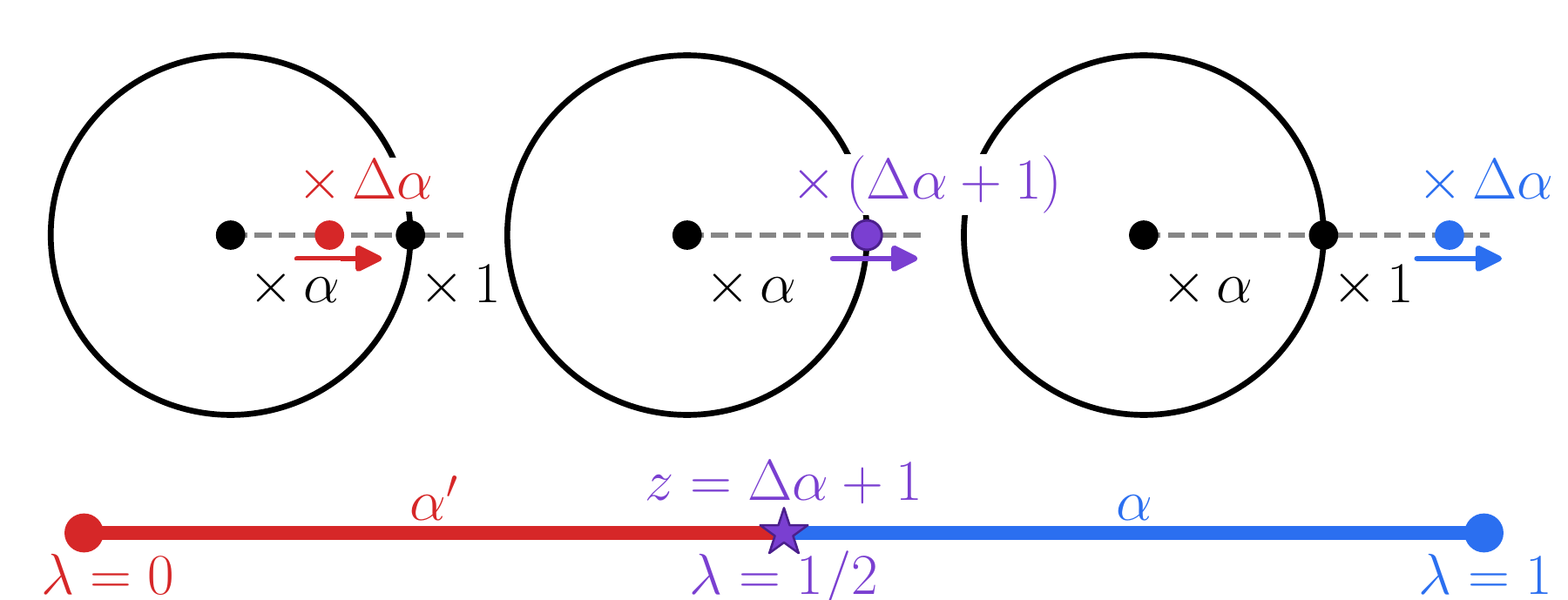}{\linewidth}{(b)}
  \end{minipage}
  \begin{minipage}[b]{1.0\linewidth}
    \panel{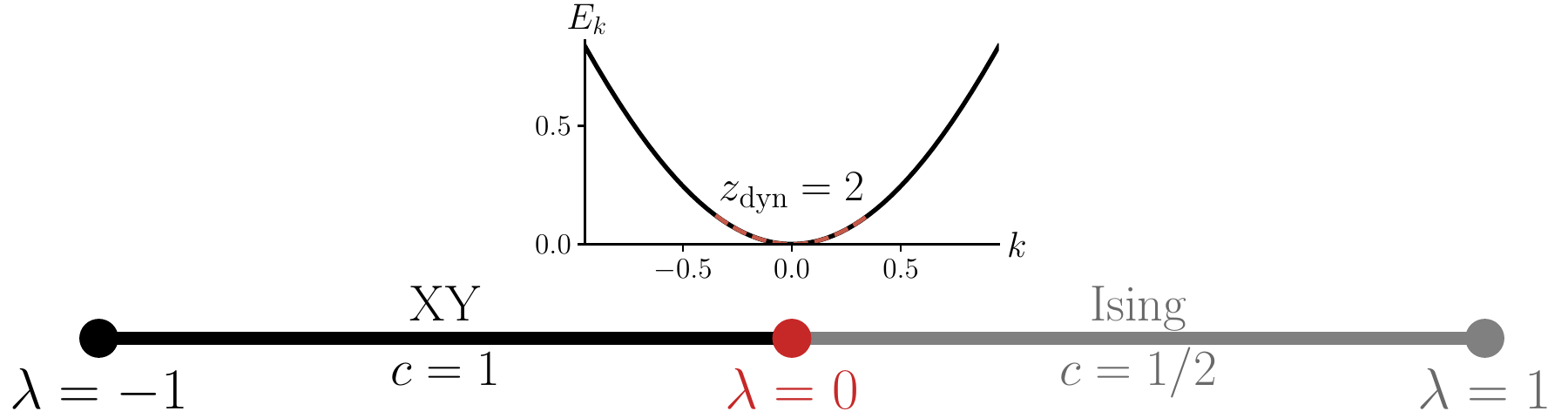}{\linewidth}{(c)}
  \end{minipage}
\caption{
Phase diagram of transitions between quantum critical lines.
(a) Representative phase diagram for a transition between two topologically distinct critical lines, illustrated by the minimal case $\alpha=0$ and $\alpha'=1$. The endpoint insets show the corresponding auxiliary-root configurations: a root on the unit circle signals criticality, while an additional root inside the unit circle encodes nontrivial topological degeneracy. The intermediate MCP exhibits $z_{\rm dyn}=2$ scaling.
(b) Auxiliary-root representation of the phase transition. In moving between the two critical lines, the roots are transferred from inside to outside the unit circle. When these roots meet the original critical root on the unit circle, the critical root becomes $(\Delta\alpha+1)$-fold degenerate, producing a Lifshitz MCP with $z_{\rm dyn}=\Delta\alpha+1$.
(c) Conventional benchmark Lifshitz MCP (see Eq.~\eqref{eq2}) between two critical lines belonging to different universality classes. The Lifshitz point separates the XY critical line with central charge $c=1$ from the Ising critical line with central charge $c=1/2$, while both sides are topologically trivial. The dispersion inset shows $z_{\rm dyn}=2$ scaling. two tuning parameters $h$ and $\gamma$ in Eq.~\eqref{eq2} are parameterized by a single parameter $\lambda$; see details of the interpolation paths in the End Matter.
}
  \label{fig:1}
\end{figure}


We now consider a single MCP between two topologically distinct critical lines labeled by $\alpha$ and $\alpha'$(Fig.~\ref{fig:1} (a)), with $\alpha' > \alpha$.  In moving from the $\alpha'$ critical line to the $\alpha$ critical line, $\Delta\alpha=\alpha'-\alpha$ zeros are taken from inside to outside the unit circle, as illustrated in Fig.~\ref{fig:1} (b). Here, we focus on the Lifshitz MCP, where these moving zeros cross the unit circle at the same momentum as the original critical zero. As a result, the zero on the unit circle becomes $(\Delta\alpha+1 )$-fold degenerate, giving a Lifshitz dynamical exponent $z=\Delta\alpha+1 > 1$. The corresponding Hamiltonian at such MCP can be constructed as
\begin{equation}
\label{eq1}
H_{\alpha,\alpha'}^{\mathrm{mc}} = \sum_j
\sum_{r=0}^{\Delta\alpha+1}
\frac{(\Delta\alpha+1)!}{r!(\Delta\alpha+1-r)!}
\left(
b_j^\dagger a_{j+\alpha+r}
+\mathrm{h.c.}
\right).
\end{equation}
In particular, the coefficients of the hopping terms follow a binomial distribution (see Supplemental Material (SM) for details of the construction). For $\alpha=0$ and $\alpha'=1$, the above Hamiltonian reduces to the MCP realized from a linear interpolation between the critical Kitaev chain ($H_0+H_1$) and the two-copy critical Kitaev chain ($H_1+H_2$)~\cite{Yu2024PRB,Choi2024PRB}. However, for the more general case $\alpha \neq \alpha'$, such a naive linear-interpolation construction breaks down and cannot realize the Lifshitz MCPs. Instead, $H_{\alpha,\alpha'}^{\mathrm{mc}}$ represents the genuine Hamiltonian describing the topologically enforced Lifshitz MCPs.

\emph{Conventional multicritical points between different universality classes.}---Before turning to a detailed discussion of topologically enforced Lifshitz MCPs, we first briefly review Lifshitz MCPs that typically emerge at the intersection of different universality classes. In fact, many previously recognized one-dimensional MCPs~\cite{Rodney2013PRB,Mohammadi2017JHEP,Temple2017SciPost,Gentle2018JHEP,Dion2021SciPost} in the literature belong to this category and will therefore be referred to as conventional MCPs in the following.

\begin{figure}[t]
  \centering
  \begin{minipage}[b]{1.0\linewidth}
    \twopanel{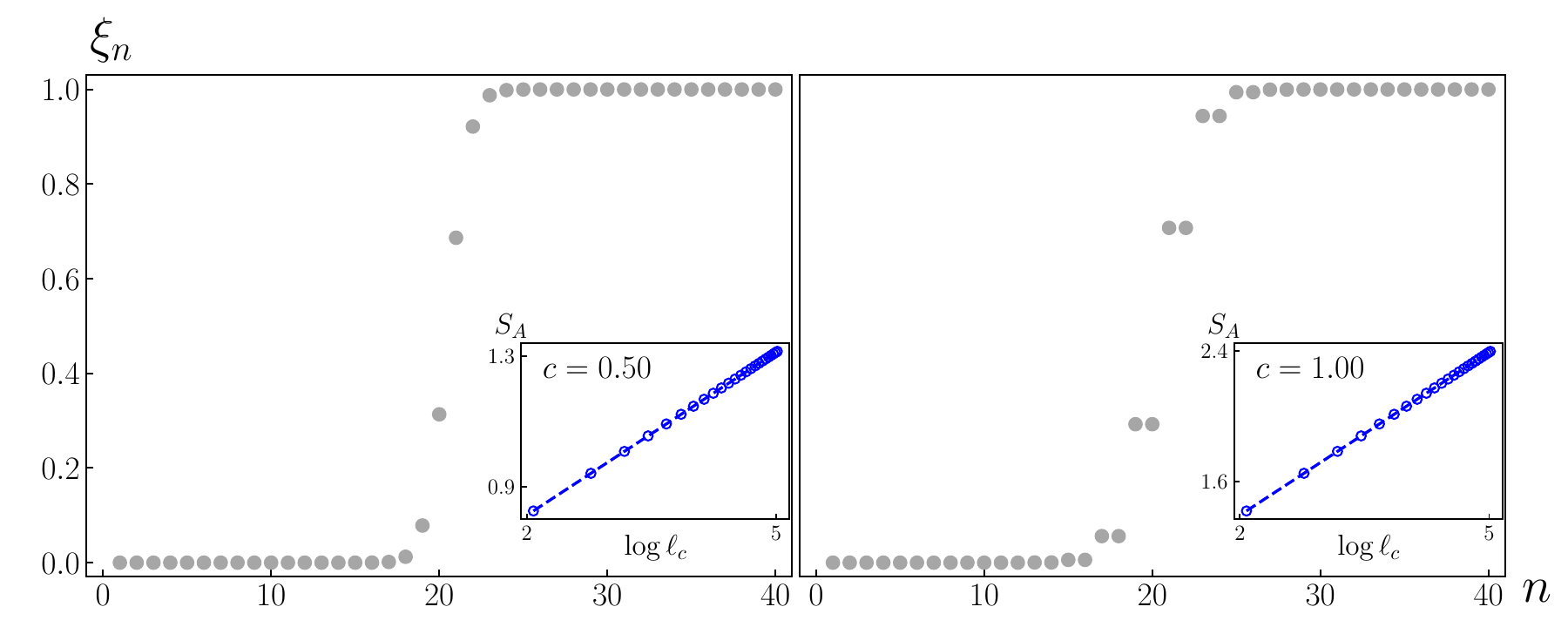}{\linewidth}{(a)}{(b)}
  \end{minipage}
  \begin{minipage}[b]{0.49\linewidth}
    \panel{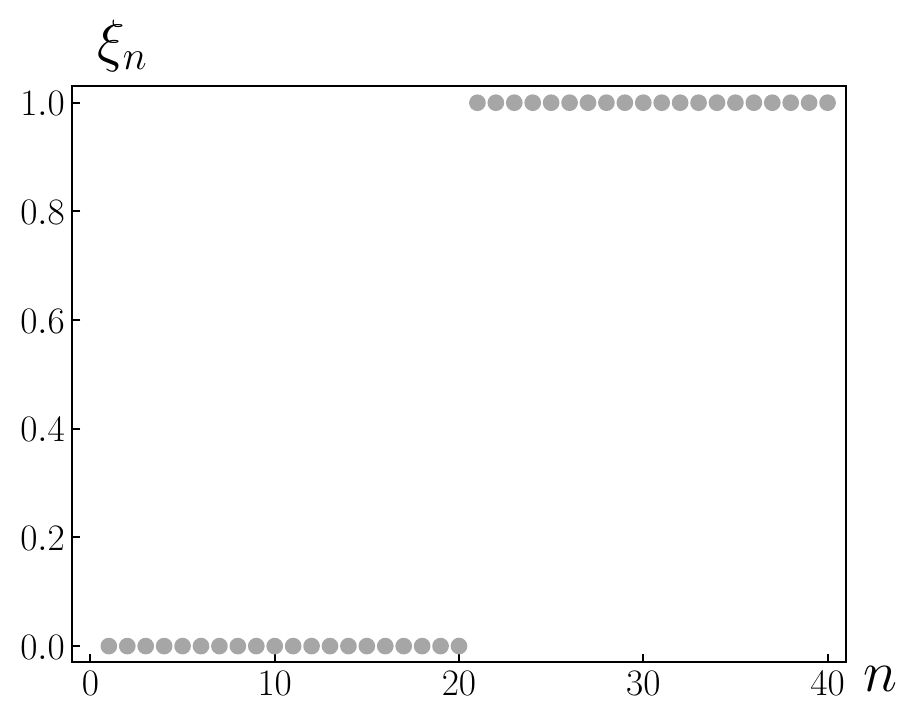}{\linewidth}{(c)}
  \end{minipage}
  \begin{minipage}[b]{0.49\linewidth}
    \panel{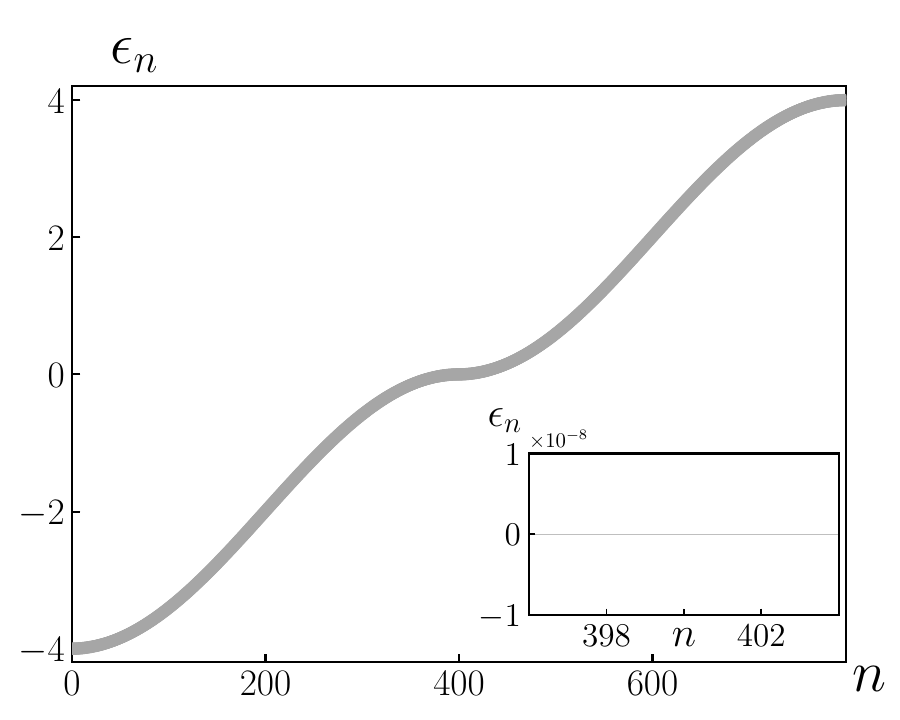}{\linewidth}{(d)}
  \end{minipage}
\caption{Bulk entanglement spectra and boundary energy spectrum near a Lifshitz MCP separating distinct universality classes. PBC entanglement spectra at two adjacent topologically trivial quantum critical points: the Ising critical point (a) and the XY critical point (b). The insets show the corresponding entanglement entropy scaling, giving central charges $c\simeq 0.50$ and $c\simeq 1.00$, respectively.
(c) PBC entanglement spectrum at the conventional Lifshitz MCP. No isolated midgap entanglement level appears.
(d) OBC single-particle energy spectrum at the same MCP. The inset zooms into an exponentially small zero-energy window, showing the absence of  boundary zero modes.
}
  \label{fig:2}
\end{figure}

To be specific, we consider the transverse-field XY chain~\cite{Zhu2006PRL,Mukherjee2007PRB} as a prototypical example illustrating the key features of conventional MCPs:
\begin{equation}
\label{eq2}
H_{\mathrm{t}\mathrm{XY}}
=
-\frac{1}{2}
\sum_j
\left[
(1+\gamma)X_jX_{j+1}
+
(1-\gamma)Y_jY_{j+1}
+
2h Z_j
\right].
\end{equation}
Here, $X_i$, $Y_i$, and $Z_i$ denote spin-$1/2$ Pauli matrices on site $i$, while $h,\gamma$ are two tuning parameters. This model is exactly solvable via Jordan--Wigner transformation (see SM Sec.~\ref{sm2} for details), and the associated phase diagram exhibits quantum critical lines belonging to different universality classes characterized by distinct central charges $c$: the Ising critical line with $c=1/2$ obtained at $h=1, \gamma\neq 0$ (see insert of Fig.~\ref{fig:2} (a)), while the XY critical line with $c=1$ is realized at $\gamma=0, |h|<1$ (see inset of Fig.~\ref{fig:2} (b)). These two critical lines meet at $(h,\gamma)=(1,0)$, where the low-energy dispersion becomes quadratic, $E(k)\sim k^2$ (see Fig.~\ref{fig:1} (c)). Moreover, the topological properties of the quantum critical lines can be diagnosed through the bulk entanglement spectrum using the recently proposed generalized Li–Haldane correspondence~\cite{guo2025generalizedlihaldanecorrespondencecritical}. As shown in Fig.~\ref{fig:2} (a) and (b), we find that the entanglement spectra of both critical lines exhibit no midgap states and are therefore topologically trivial. We further calculate the entanglement [Fig.~\ref{fig:2} (c)] and energy spectrum [Fig.~\ref{fig:2} (d)] at the MCP under periodic (PBCs) and open boundary conditions (OBCs), respectively. Neither spectrum exhibits nontrivial topological degeneracy, indicating that the conventional MCP is topologically trivial. More generally, conventional Lifshitz MCPs formed by topologically
trivial critical lines are expected to retain the same topologically
trivial character: since the neighboring critical theories carry no
protected entanglement-spectrum or boundary degeneracies, their meeting
does not require such degeneracies to appear. Thus, such conventional
MCPs are topologically trivial and arise from changes in the universality
classes of critical lines.

\emph{Multicritical points between topologically distinct quantum critical lines.}---Besides the most common mechanism for realizing MCPs discussed above, we now turn to a novel class of Lifshitz MCPs realized through an alternative mechanism enforced solely by changes in the topology of neighboring critical lines that nevertheless share the same central charge.

\begin{figure}[t]
  \centering
  \begin{minipage}[b]{1.0\linewidth}
    \twopanel{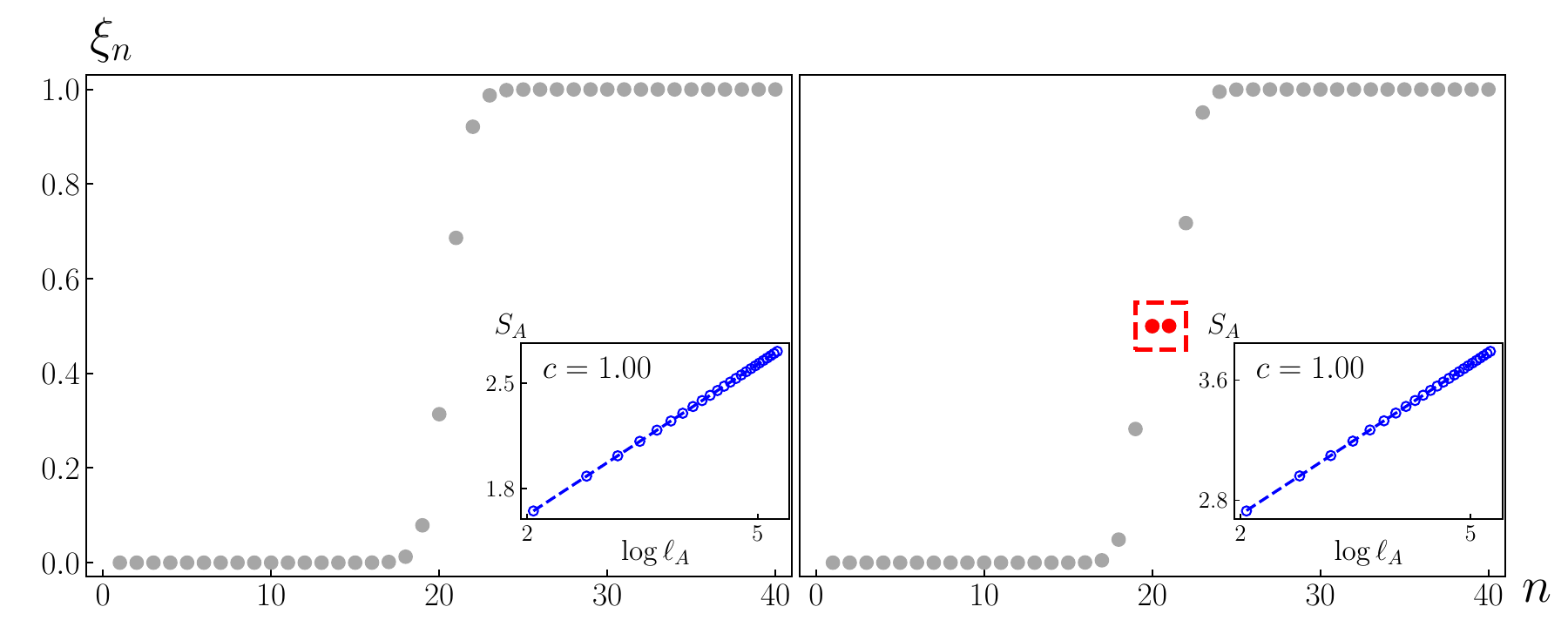}{\linewidth}{(a)}{(b)}
  \end{minipage}
  \begin{minipage}[b]{0.49\linewidth}
    \panel{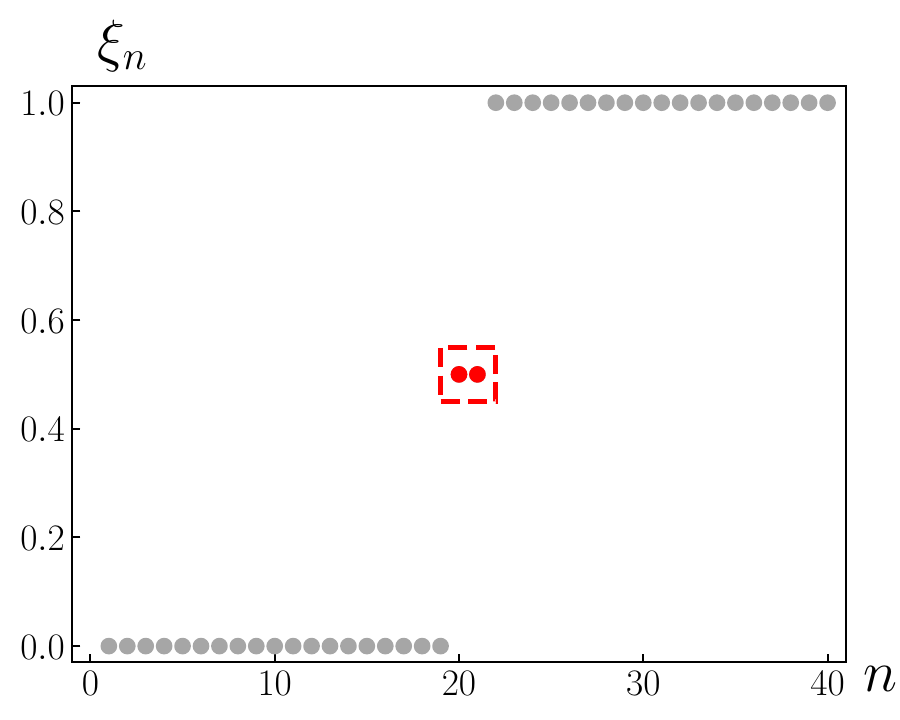}{\linewidth}{(c)}
  \end{minipage}
  \begin{minipage}[b]{0.49\linewidth}
    \panel{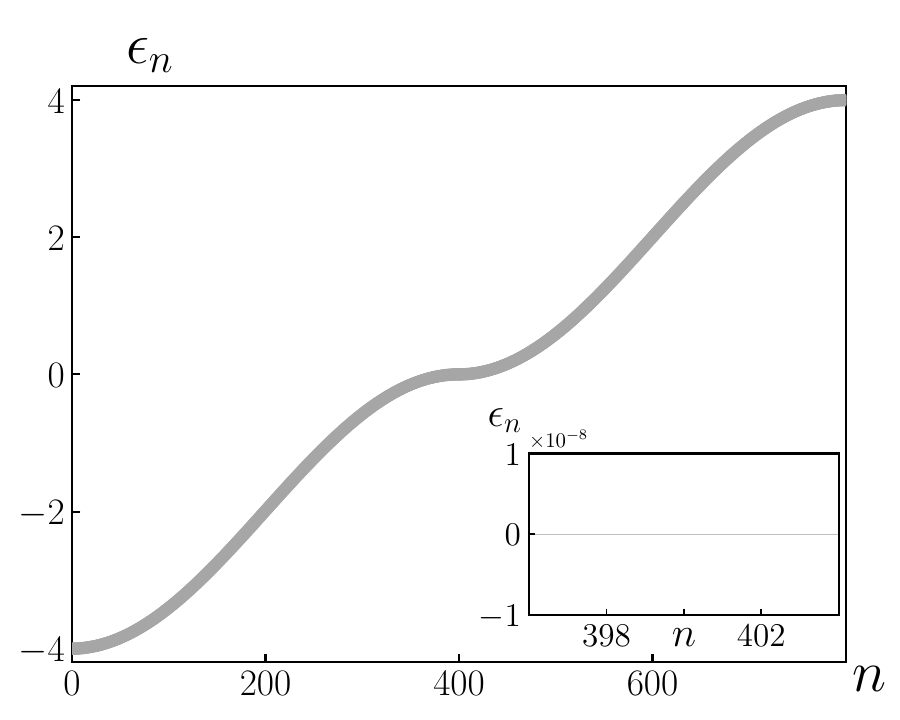}{\linewidth}{(d)}
  \end{minipage}
  \caption{
Bulk entanglement spectra and physical boundary spectrum near a topologically enforced Lifshitz MCP. PBC entanglement spectra at two adjacent quantum critical points: a topologically trivial critical point (a) and a topologically nontrivial critical point (b). The insets show the corresponding entanglement entropy scaling, both giving the same central charge $c\simeq 1$.
(c) PBC entanglement spectrum at the topologically enforced Lifshitz MCP, where midgap entanglement levels appear.
(d) OBC energy spectrum at the same MCP. 
The inset zooms into a small window around zero energy and shows no zero-energy modes. 
Thus, although the PBC entanglement spectrum develops midgap levels, these levels are not reflected as protected zero modes in the physical OBC spectrum. 
This reveals a breakdown of the naive Li--Haldane correspondence at such topologically enforced multicriticality.
}
  \label{fig:3}
\end{figure}


We choose the case $\alpha=0$ and $\alpha'=1$ in Eq.~\eqref{eq1} as an illustrative example. The corresponding Hamiltonian describes an MCP between the critical Kitaev chain and the two-copy critical Kitaev chain. In sharp contrast to conventional MCPs, the adjacent critical lines belong to the same universality class (see SM Sec.~\ref{sm:alpha_critical_chain} for further evidences) and are both characterized by  $c=1$, yet possess distinct topological properties: the two-copy Kitaev critical line hosts a twofold-degenerate midgap state in the bulk entanglement spectrum, whereas the Kitaev critical line does not, as shown in Fig.~\ref{fig:3} (a) and (b). This implies that a phase transition must occur between these topologically distinct critical lines, and the corresponding transition point—a Lifshitz MCP with $z_{\mathrm{dyn}}=2$—is enforced solely by the change in topology along the critical lines, as confirmed in Fig.~\ref{fig:1} (a). Similarly, we further calculate the bulk entanglement spectrum at this topologically enforced MCP [Fig.~\ref{fig:3} (c)] and find a robust twofold-degenerate midgap state, indicating that the MCP itself is topologically nontrivial. However, upon calculating the energy spectrum under OBCs [Fig.~\ref{fig:3} (d)], we surprisingly find the absence of degenerate zero energy modes, implying instead a topologically trivial boundary energy spectrum. One might then wonder whether the midgap modes in the entanglement spectrum are accidental rather than topological. For the even-$z$ Lifshitz MCP considered here, however, the winding number is still well defined and guarantees the robust topology of the midgap entanglement structure; see the detailed discussion in SM Sec.~\ref{sm4}. This directly signals a breakdown of the Li–Haldane correspondence~\cite{Li2008PRL}—an alternative formulation of the bulk–boundary correspondence that constitutes one of the fundamental principles of topological physics. In fact, the central contradiction is that the number of topological degeneracies encoded in the bulk entanglement spectrum and the energy spectrum under OBCs are inequivalent ($2\neq0$ in the present case; see SM Sec.~\ref{sm4} for the general $\alpha,\alpha'$ case). Moreover, in SM Sec.~\ref{sm3}, we further show that this breakdown at the topologically enforced multicriticality remains robust against moderate symmetry-preserving disorder and interaction.

\begin{figure}[t]
  \raggedright
  \begin{minipage}[b]{1.0\linewidth}
    \raggedright
    \makebox[\linewidth][l]{\panelH{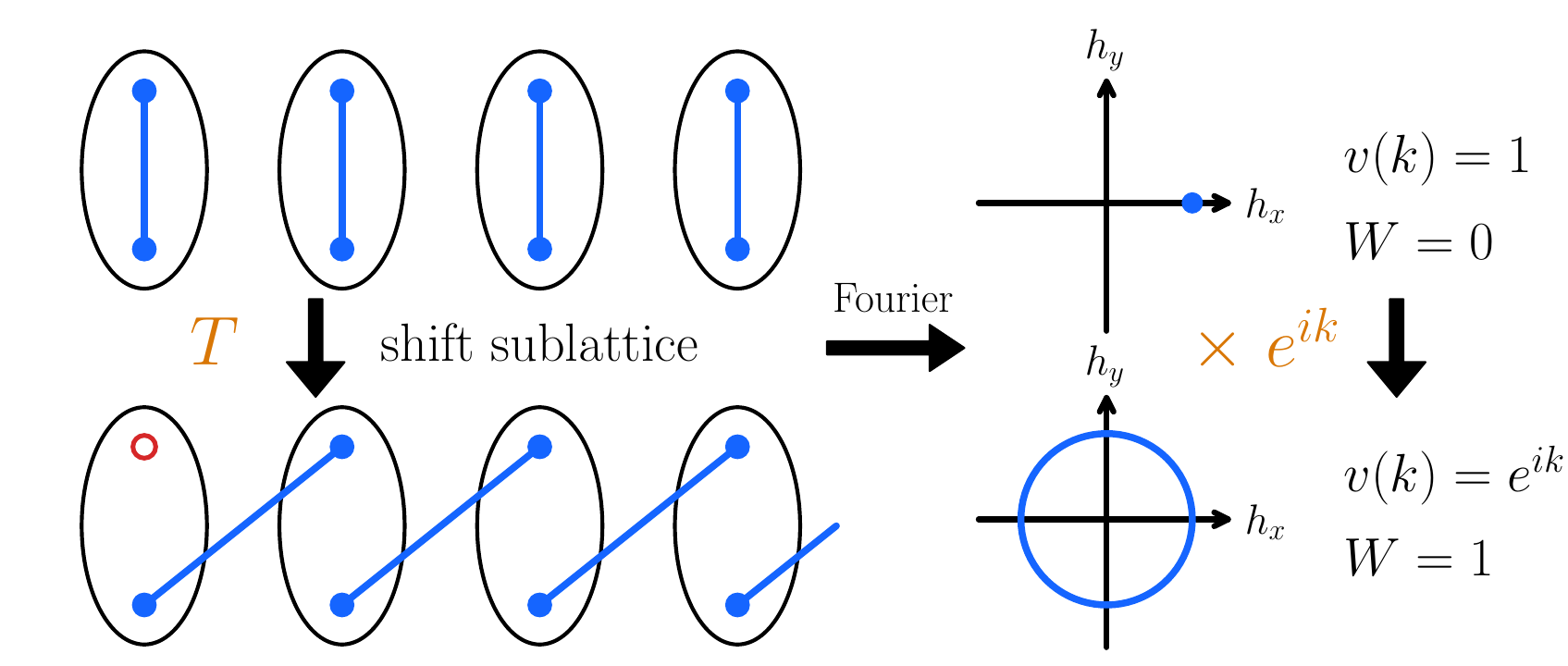}{3.3cm}{(a)}}
  \end{minipage}

  \begin{minipage}[b]{1.0\linewidth}
    \raggedright
    \makebox[\linewidth][l]{\panel{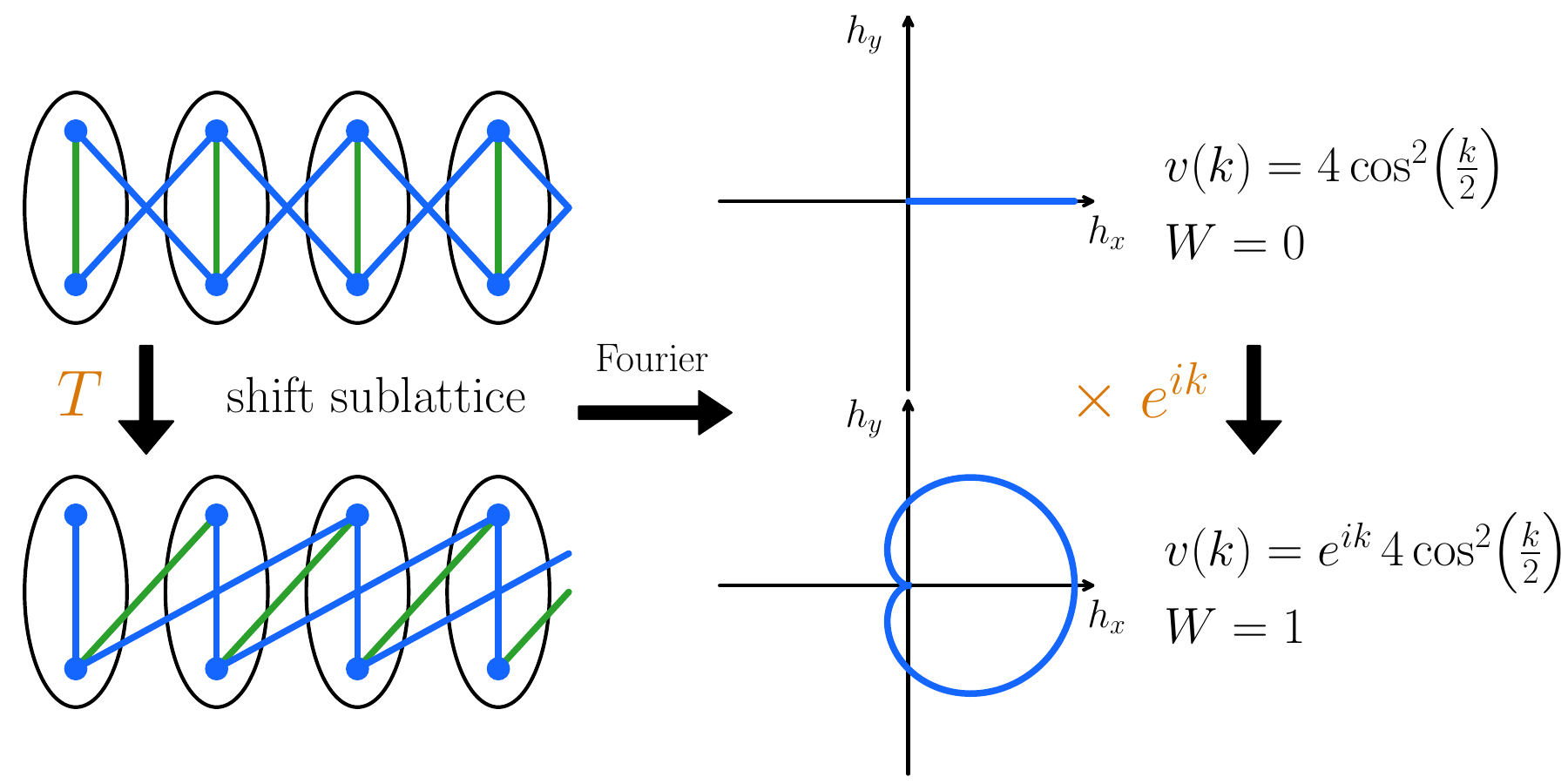}{\linewidth}{(b)}}
  \end{minipage}
  \caption{RG-limit picture for the breakdown of the Li--Haldane correspondence.
Each oval denotes one physical unit cell, containing an $A_j$ site on the
upper row and a $B_j$ site on the lower row. The map $T$ denotes a relative
translation of the $B$ sublattice, $B_j\to B_{j-1}$, which multiplies the
off-diagonal Bloch element $v_k$ by $e^{ik}$ under PBC and thereby increases the
winding number by one. The colored bonds indicate different hopping amplitudes
in the real-space coupling pattern, with the green bonds having twice the
amplitude of the blue bonds.
(a) In a gapped SPT chain, each site couples to only one partner: shifting the
sublattice changes \(v(k)=1\) to \(v(k)=e^{ik}\), and under OBC this shift changes the boundary termination, leaving an uncoupled boundary site that forms an edge mode (red circle).
(b) At the \(z_{\mathrm{dyn}}=2\) Lifshitz MCP, the coupling has finite
spatial width. Although the same shift changes
\(v(k)=4\cos^2(k/2)\) to \(v(k)=e^{ik}4\cos^2(k/2)\) and hence gives
\(W=1\), it does not isolate a boundary site in real space. Therefore the PBC
entanglement-spectrum mid-gap mode is not accompanied by a corresponding
protected OBC edge mode, explaining the failure of the naive Li--Haldane
correspondence.
}
  \label{fig:4}
\end{figure}

\emph{Breakdown of the Li–Haldane correspondence at topologically enforced multicriticality.}---We now provide a simple physical picture for why the Li–Haldane correspondence breaks down at topologically nontrivial MCPs, while leaving the detailed discussion to SM Sec.~\ref{sm4}. Specifically, we consider the low-energy theory of chiral-symmetric free-fermion models in the renormalization-group (RG) limit. For gapped phases, the trivial and topological insulator can be related by a relative shift of one sublattice, which cannot be smoothly implemented in a local and symmetry-preserving way. Such an operation generates a nontrivial winding number $W$ of the off-diagonal block $v_k$, which directly corresponds to the $2|W|$-fold degenerate midgap states in the entanglement spectrum from the perspective of the correlation matrix~\cite{Peschel_2009,Chang2020PRR}. Simultaneously, the relative shift operation changes the boundary termination and leaves behind dangling boundary sites whose number is determined by the relative translation distance, thereby giving rise to topological edge modes under OBCs [see Fig.~\ref{fig:4} (a)]. This provides a simple physical understanding of the well-established Li–Haldane correspondence in gapped topological insulators~\cite{Fidkowski2010PRL}. Similarly, for the topologically nontrivial MCPs with even $z$ considered here, although one may start from a topologically trivial MCP and perform a relative shift operation to generate a nontrivial topological winding number, the boundary sites remain coupled because the higher-order derivative terms in the Lifshitz low-energy theory have a finite spatial width, as illustrated in Fig.~\ref{fig:4} (b).  This is the key reason why the Li–Haldane correspondence breaks down at topologically nontrivial MCPs. In general, since the Li--Haldane correspondence relates physical boundary modes to the boundary modes induced by an entanglement cut, it is sensitive to boundary physics; thus, the finite width of the higher-order derivative coupling in the Lifshitz theory can modify the boundary behavior and make the usual correspondence no longer guaranteed. Such a mechanism cannot occur in previously known gapped or gapless topological phases, since their low-energy theories contain at most first-order derivative structures and therefore do not possess the finite spatial width needed to produce this boundary mismatch. For odd-$z$ Lifshitz MCPs, a similar mismatch between physical OBC edge modes and entanglement-spectrum zero modes can occur; further details are given in SM Sec.~\ref{sm4}. Although the above physical picture is formulated in terms of decoupled boundary modes in the RG limit, the argument remains valid even when the boundary couplings are finite but RG irrelevant. 

In SM Sec.~\ref {sm4}, we provide a general criterion for the emergence of nontrivial topology in the Lifshitz MCPs described by Eq.~\eqref{eq1}, which is briefly summarized below. For an even-$z$ Lifshitz MCP connecting the $\alpha$ and $\alpha'$ critical lines, the winding number is $W=(\alpha+\alpha'+1)/2$, and the nontrivial topology of such an MCP originates from the imbalanced hopping structure of the adjacent critical lines. By contrast, for the balanced case $\alpha+\alpha'=-1$ (e.g., $\alpha=0$ and $\alpha'=-1$), the adjacent critical lines have hopping structures $H_0+H_1$ and $H_{-1}+H_0$, and the topologically enforced MCP remains topologically trivial and can be smoothly connected to a conventional MCP. 
We further emphasize two important remarks. First, the minimal model realizing a topologically nontrivial MCP corresponds to the transition between topologically trivial ($\alpha=0$) and nontrivial ($\alpha'=1$) quantum critical lines. This is fundamentally different from the realization of topologically nontrivial quantum critical points, which cannot emerge from transitions between topologically trivial and nontrivial \emph{insulating} phases~\cite{Verresen2018PRL,verresen2020topologyedgestatessurvive}. Second, the topologically enforced transitions revealed here provide a unified framework for understanding many unconventional one-dimensional MCPs reported in previous literature~\cite{Malard2020PRR,Malard2020PRB,Kumar2021SR,Kumar2023PRB,kumar2025topologicallynontrivialmulticriticalpoints,prembabu2025multicriticalitypurelygaplessspt}.






\emph{Concluding remarks.}---To summarize, we systematically construct a novel class of one-dimensional Lifshitz MCPs that are qualitatively distinct from conventional ones. Specifically, such multicriticality is enforced solely by changes in the topology of neighboring quantum critical lines and becomes topologically nontrivial in the sense that it exhibits robust degenerate midgap states in the bulk entanglement spectrum when the adjacent critical lines possess imbalanced hopping structures. By contrast, MCPs between critical lines with balanced hopping remain topologically trivial and can be smoothly connected to conventional MCPs. More importantly, we unambiguously demonstrate that the Li–Haldane correspondence surprisingly breaks down at these topologically nontrivial MCPs—a phenomenon unique to topological multicriticality that can be understood through a simple physical picture.

In the future, it would be particularly interesting to explore topologically enforced MCPs in higher-dimensional systems, where both topological and multicritical phenomena are expected to be substantially richer than in one dimension. It is also worthwhile to develop a finite-size scaling theory tailored to these novel MCPs. From an experimental perspective, the one-dimensional free-fermion models considered here have been recently realized in phononic systems~\cite{Liu2023PRAp,Xiong2025PRL} and electronic circuits~\cite{Dong2021PRR,Dias2022PRB}, where multiple long-range hopping processes can be conveniently engineered and controlled within the same settings.

\textit{Acknowledgement}: X.-J. Yu was supported by the National Natural Science Foundation of China (Grant No.12405034) and a start-up grant from Eastern Institute of Technology, Ningbo. K.-H. C. was supported by the National Science and Technology Council of Taiwan under Grant No. NSTC 114-2112-M-007-015.

\bibliographystyle{apsrev4-2}
\let\oldaddcontentsline\addcontentsline
\renewcommand{\addcontentsline}[3]{}
\bibliography{main.bib}

\appendix
\section{\large{End Matter}}
\twocolumngrid

\textit{Physical observables for diagnosing criticality and topology}---We first introduce the entanglement spectrum diagnostic used in the main text.
In free-fermion topological systems, a spatial entanglement cut can be viewed as
creating virtual boundaries inside the bulk wave function. If the bulk state carries
nontrivial topology, these virtual boundaries can host protected modes, which appear
as robust midgap levels pinned at $\xi=1/2$ in the correlation spectrum. Therefore,
the presence or absence of stable $\xi=1/2$ levels provides a bulk diagnostic of
topological structure. This is the free-fermion version of the Li--Haldane
correspondence: the entanglement cut diagnoses the boundary structure encoded in the
ground state, even without imposing a physical open boundary.

For the complex-fermion lattice with two sublattices in each unit cell, we denote the
fermion operators on the $A$ and $B$ sublattices by $a_j$ and $b_j$, respectively.
For a subsystem $\mathcal A$ containing several unit cells, the restricted correlation
matrix is written in the sublattice basis. Its $2\times2$ block between unit cells
$i$ and $j$ is
\begin{equation}
    C_{ij}
    =
    \begin{pmatrix}
        \langle a_i^\dagger a_j\rangle
        &
        \langle a_i^\dagger b_j\rangle
        \\[1mm]
        \langle b_i^\dagger a_j\rangle
        &
        \langle b_i^\dagger b_j\rangle
    \end{pmatrix},
    \qquad i,j\in \mathcal A .
\end{equation}
The full restricted correlation matrix $C_{\mathcal A}$ is obtained by assembling
these $2\times2$ blocks over all unit cells inside the subsystem. Its eigenvalues
$C_{\mathcal A}\varphi_n=\xi_n\varphi_n$, with $0\leq \xi_n\leq1$, define the
correlation spectrum plotted in the main text.

For the Majorana-chain representation, we combine two Majorana operators into one
complex fermion by
\begin{equation}
    \gamma_{2j-1}=c_j+c_j^\dagger,
    \qquad
    \gamma_{2j}=-i(c_j-c_j^\dagger).
\end{equation}
Since particle number is not generally conserved in this representation, both normal
and anomalous contractions are included. In the Nambu basis, the $2\times2$ block
between sites $i$ and $j$ is
\begin{equation}
    C^{\rm Nambu}_{ij}
    =
    \begin{pmatrix}
        \langle c_i^\dagger c_j\rangle
        &
        \langle c_i^\dagger c_j^\dagger\rangle
        \\[1mm]
        \langle c_i c_j\rangle
        &
        \langle c_i c_j^\dagger\rangle
    \end{pmatrix},
    \qquad i,j\in \mathcal A .
\end{equation}
The full restricted Nambu correlation matrix $C_{\mathcal A}^{\rm Nambu}$ is obtained
by assembling these blocks over all sites inside the subsystem. Its eigenvalues come
in particle-hole-related pairs $\xi_\ell$ and $1-\xi_\ell$. In both formulations,
robust midgap levels at $\xi=1/2$ are used as the correlation-spectrum signature of
topological virtual boundary modes.

The entanglement entropy shown in the insets is computed from the same correlation
data and is used as a supplementary diagnostic for identifying the central charge of
the adjacent critical lines. For the number-conserving complex-fermion case,
\begin{align}
    S_A
    &=
    -\mathrm{Tr}\left[
        C_A\ln C_A+
        (1-C_A)\ln(1-C_A)
    \right]\nonumber \\
    &=
    -\sum_n
    \left[
        \xi_n\ln\xi_n+
        (1-\xi_n)\ln(1-\xi_n)
    \right],
\end{align}
For the Nambu/Majorana formulation, the same expression applies after keeping one
eigenvalue from each particle-hole-related pair, or equivalently with an additional
factor $1/2$ in front of the trace over $C_{\mathcal A}^{\rm Nambu}$.

\textit{Brief review of the topological physics in critical free-fermion systems.}---To illustrate the topological physics of quantum critical systems more intuitively, we consider the $\alpha$ chain in the Majorana representation~\cite{Verresen2018PRL}:
\begin{equation}
H_{\alpha}^{\mathrm{Maj}} = i \sum_{n} \gamma_{2n} \gamma_{2(n+\alpha)-1},
\end{equation}
where $\gamma_{2n-1}$ and $\gamma_{2n}$ denote two Majorana fermion species on each site. The Hamiltonian $H_{\alpha}^{\mathrm{Maj}}$ preserves particle-hole and time-reversal symmetries and belongs to the BDI symmetry class. In the complex-fermion representation, this model can be mapped onto an extended Su--Schrieffer--Heeger model with $\alpha$-range hopping in symmetry class AIII, which is the formulation adopted in the main text. In general, $H_{\alpha}^{\mathrm{Maj}}$ supports $|\alpha|$ Majorana zero modes at each edge and can therefore be viewed as a stack of $|\alpha|$ Kitaev chains. Consequently, its ground state realizes a gapped topological phase with a $2^{|\alpha|}$-fold edge degeneracy. Topological critical Hamiltonians can then be constructed from linear combinations of different $\alpha$ chains, giving rise to continuous phase transitions between gapped phases with distinct topological invariants. Importantly, whenever such a transition involves a change in a nonzero topological index, the resulting quantum critical point hosts exponentially localized Majorana edge modes~\cite{Verresen2018PRL,verresen2020topologyedgestatessurvive}. This provides a general guiding principle for engineering gapless symmetry-protected topological (SPT) phases in free-fermion systems across arbitrary spatial dimensions. Moreover, we emphasize that conventional topological invariants, such as the winding number, become ill-defined at topologically nontrivial critical points because the winding structure is intrinsically singular at the origin of parameter space. Instead, recent progress~\cite{guo2025generalizedlihaldanecorrespondencecritical} has established that the bulk entanglement spectrum provides an alternative diagnostic to identify nontrivial topology in free-fermion critical systems in arbitrary dimensions.

On the other hand, the gapped and gapless SPT physics of the $\alpha$ chain can be
understood through the root-counting description based on the complex function
$v(\beta)$~\cite{Verresen2018PRL,Jones2019JSP}, where $\beta=e^{ik}$ is the
complex variable associated with the Bloch factor. With the convention used in the
main text,
\begin{equation}
    H(k)=
    \begin{pmatrix}
        0 & v_k^*\\
        v_k & 0
    \end{pmatrix},
\end{equation}
a factor $e^{ikr}$ in $v_k$ corresponds to a real-space hopping between the $B$ sublattice in unit cell $j$ and the $A$ sublattice in unit cell $j+r$.

We first consider the finite-polynomial case,
\begin{equation}
    v(\beta)=\sum_{r\geq0}t_r\beta^r ,
\end{equation}
for which the root-counting interpretation under OBC is most transparent. The opposite
orientation, described by a finite polynomial in $\beta^{-1}$, can be treated
analogously and gives edge modes with the opposite chirality.

A simple example is the mass-inversion form
\begin{equation}
    v(\beta)=\beta-\kappa,
\end{equation}
with $\kappa = 1+m$ and $m$ is the mass of the system. In real space, this corresponds to
\begin{equation}
    H
    =
    \sum_{j\geq0}
    \left[
        b_j^\dagger a_{j+1}
        -
        \kappa\,b_j^\dagger a_j
        +{\rm h.c.}
    \right].
\end{equation}
The single-particle OBC equations $H\ket{\psi}=E\ket{\psi}$ are
\begin{align}
    \psi_{A,j+1}-\kappa\psi_{A,j}
    &=
    E\psi_{B,j},
    \qquad j\geq0,\label{eq:recur_a}\\
    -\kappa\psi_{B,0}
    &=
    E\psi_{A,0},\label{eq:boundary}\\
    \psi_{B,j-1}-\kappa\psi_{B,j}
    &=
    E\psi_{A,j},
    \qquad j\geq1 .\label{eq:recur_b}
\end{align}
We now focus on zero modes, $E=0$. Due to chiral symmetry, the zero-energy
eigenvalue equations decouple into two sublattice sectors. Since
Eqs.~\eqref{eq:recur_a} and~\eqref{eq:recur_b} are constant-coefficient recursion
relations, their elementary solutions can be written in the geometric form
\begin{equation}
    \psi_{A/B,j}
    =
    \beta^j\psi_{A/B},
    \qquad j\geq0 .
\end{equation}
The boundary equation~\eqref{eq:boundary} gives $-\kappa\psi_B=0$. For $\kappa\neq0$, this
implies $\psi_B=0$, and Eq.~\eqref{eq:recur_b} then consistently gives
$\psi_{B,j}=0$ for all $j\geq0$. Thus, for the left edge, the zero-mode problem
reduces to the $A$-sublattice sector.

Substituting $\psi_{A,j}=\beta^j\psi_A$ into Eq.~\eqref{eq:recur_a} at $E=0$ gives
\begin{equation}
    \beta^j(\beta-\kappa)\psi_A=0,
    \qquad j\geq0 .
\end{equation}
For a nontrivial solution with $\psi_A\neq0$, we obtain
\begin{equation}
    v(\beta)=\beta-\kappa=0 .
\end{equation}
Thus the root $\beta=\kappa$ determines the spatial profile
$\psi_{A,j}\propto \beta^j$. For $|\beta|<1$, this solution is normalizable and
localized near the left boundary, corresponding to a topological edge mode. For
$|\beta|=1$, the solution is non-decaying and represents a bulk zero mode, indicating
a bulk gap closing. For $|\beta|>1$, the solution grows away from the boundary and is
not normalizable; the system is therefore trivial.

This example actually illustrates the general structure for the finite-polynomial case, where $v(\beta)$ contains only nonnegative powers of $\beta$: the zero-energy boundary equations select one chiral sector, and the remaining recursion yields the root condition $v(\beta)=0$.

The $\alpha$ critical chain used in the main text is constructed from two neighboring
hopping Hamiltonians, $H_\alpha$ and $H_{\alpha+1}$, where
\begin{equation}
    H_\alpha=-\sum_j\left(b_j^\dagger a_{j+\alpha}+{\rm h.c.}\right).
\end{equation}
With the convention above, $H_\alpha$ contributes a factor $\beta^\alpha$ to the off-diagonal
Bloch element, up to an overall sign. Therefore the critical chain obtained by
combining $H_\alpha$ and $H_{\alpha+1}$ has
\begin{equation}
    v_\alpha(\beta)\propto
    \beta^\alpha+\beta^{\alpha+1}
    =
    \beta^\alpha(\beta+1).
\end{equation}
The factor $\beta^\alpha$ gives $\alpha$ roots at the origin and therefore encodes
the $\alpha$ topological edge modes. The factor $(\beta+1)$ gives a root at
$\beta=-1$, which lies on the unit circle and therefore signals the bulk critical
point.

The above examples illustrate the criterion used in the main text. For a finite
polynomial $v(\beta)$ with nonnegative powers of $\beta$, roots of $v(\beta)=0$ inside
the unit circle give normalizable topological edge modes, while roots on the unit
circle correspond to bulk gap closing.

For a more general finite-range hopping problem with both directions present,
$v(\beta)$ can contain both positive and negative powers of $\beta$,
\begin{equation}
    v(\beta)=\sum_{r=r_{\rm min}}^{r_{\rm max}}t_r\beta^r .
\end{equation}
The roots of $v(\beta)=0$ still determine the elementary bulk-recursion solutions,
but the OBC boundary equations must now also be imposed. For the left
$A$-sublattice zero-mode equation $D_{\rm OBC}\psi_A=0$, the OBC termination removes
all terms with $j+r<0$. Thus, when $r_{\rm min}<0$, the first boundary equations are
\begin{equation}
    \sum_{r=-j}^{r_{\rm max}}
    t_r\,\psi_{A,j+r}=0,
    \qquad
    j=0,1,\ldots,-r_{\rm min}-1 .
\end{equation}
These are the $-r_{\rm min}$ boundary constraints. For
$j\geq -r_{\rm min}$, all hopping terms are present and the equation becomes the
full bulk recursion,
\begin{equation}
    \sum_{r=r_{\rm min}}^{r_{\rm max}}
    t_r\,\psi_{A,j+r}=0 .
\end{equation}
Using the geometric form $\psi_{A,j}=\beta^j\psi_A$, the bulk recursion gives
\begin{equation}
    v(\beta)=0 .
\end{equation}

Equivalently, multiplying by $\beta^{-r_{\rm min}}$, one obtains an ordinary
polynomial,
\begin{equation}
    \beta^{-r_{\rm min}}v(\beta)=0 .
\end{equation}
The roots of this equation with $|\beta|<1$ give the left-decaying bulk solutions.
If a root has multiplicity $m$, it contributes $m$ independent solutions. The OBC
boundary equations then impose homogeneous linear constraints on the coefficients of
these decaying solutions. Therefore, for the hard-wall termination considered here,
the number of left $A$-sublattice edge modes is
\begin{equation}
    N_L^A
    =
    N_<\!\left[\beta^{-r_{\rm min}}v(\beta)\right]
    -
    \max(0,-r_{\rm min}),
\end{equation}
when the right-hand side is positive; otherwise no left $A$-sublattice zero mode
survives. Here $N_<$ counts roots inside the unit circle, with multiplicity. The
subtraction has a simple linear-algebra meaning: if the boundary constraints remove
all independent decaying solutions, the only remaining solution is the trivial one.

The same reasoning determines the other boundary sectors. The right-edge partner of
a left $A$-sublattice zero mode lies in the $B$-sublattice sector, while the
right-edge partner of a left $B$-sublattice zero mode lies in the $A$-sublattice
sector. Thus it is sufficient to state the counting rule for the left boundary; the
right-boundary modes follow by reversing the boundary orientation and exchanging the
sublattice sector. For the left $B$-sublattice sector, the zero-mode equation is
$D_{\rm OBC}^\dagger\psi_B=0$, which reverses the hopping orientation. Thus the
left-boundary constraints are controlled by $r_{\rm max}$ rather than
$r_{\rm min}$.

We illustrate this rule using the interpolation between the $\alpha=-2$ and
$\alpha'=1$ critical chains. The root-transfer path is
\begin{equation}
    v_\lambda(\beta)
    =
    \beta^{-2}(\beta+1)
    \left[(1-\lambda)\beta+\lambda\right]^3,
    \qquad 0\leq\lambda\leq1 .
\end{equation}
At $\lambda=0$, this reduces to
\begin{equation}
    v_\lambda(\beta)=\beta(\beta+1),
\end{equation}
corresponding to the $\alpha'=1$ critical line. At $\lambda=1$, it reduces to
\begin{equation}
    v_\lambda(\beta)=\beta^{-2}(\beta+1),
\end{equation}
corresponding to the $\alpha=-2$ critical line.

We first analyze the left $A$-sublattice zero-mode equation. For this interpolation,
$r_{\rm min}=-2$, so there are two left-boundary constraints in the $A$ sector.
Multiplying by $\beta^2$, we find
\begin{equation}
    \beta^2 v_\lambda(\beta)
    =
    (\beta+1)
    \left[(1-\lambda)\beta+\lambda\right]^3 .
\end{equation}
The root $\beta=-1$ lies on the unit circle and corresponds to the critical bulk
root. The remaining three roots are degenerate at
\begin{equation}
    \beta_\lambda
    =
    -\frac{\lambda}{1-\lambda}.
\end{equation}
For $\lambda<1/2$, these three roots lie inside the unit circle and give three
left-decaying candidate solutions. Since the root $\beta_\lambda$ has multiplicity
three, the most general left-decaying solution generated by these roots is
\begin{equation}
    \psi_{A,j}
    =
    \left(C_0+C_1j+C_2j^2\right)\beta_\lambda^j .
\end{equation}
The two boundary constraints determine two independent relations among
$C_0,C_1,C_2$, leaving one nontrivial edge-mode solution. Explicitly, solving the
two boundary equations gives
\begin{equation}
    (C_0,C_1,C_2)\propto (2,3,1),
\end{equation}
and therefore the surviving left-edge zero mode can be written as
\begin{equation}
    \psi_{A,j}
    =
    \mathcal N\,(j+1)(j+2)\,\beta_\lambda^j,
    \qquad
    \psi_{B,j}=0,
\end{equation}
where $\mathcal N$ is a normalization constant. Hence only one independent
normalizable left $A$-sublattice zero mode remains,
\begin{equation}
    N_L^A=3-2=1 .
\end{equation}
For $\lambda>1/2$, the three moving roots lie outside the unit circle, so there is
no left $A$-sublattice edge mode.

We now analyze the left $B$-sublattice zero-mode equation
$D_{\rm OBC}^\dagger\psi_B=0$. In this sector the relevant Bloch element is the
complex-conjugated off-diagonal block. We denote the corresponding characteristic
function by $\bar v(\beta)$, obtained from $\bar v_k=v_k^*$ and then writing
$e^{ik}$ as $\beta$. We denote the corresponding
characteristic function by
\begin{equation}
    \bar v(\beta)
    =
    \sum_{r=r_{\rm min}}^{r_{\rm max}}
    t_r^* \beta^{-r}.
\end{equation}
In this sector the left-boundary constraints are controlled by $r_{\rm max}$ of the
original $v_\lambda(\beta)$. Since $r_{\rm max}=2$, the left boundary gives two
boundary constraints. Multiplying by $\beta^2$, we obtain
\begin{equation}
    \beta^2\bar v_\lambda(\beta)
    =
    (\beta+1)
    \left[\lambda\beta+(1-\lambda)\right]^3 .
\end{equation}
Again, $\beta=-1$ is the critical bulk root. The three moving roots are now
\begin{equation}
    \beta_\lambda
    =
    -\frac{1-\lambda}{\lambda}.
\end{equation}
For $\lambda>1/2$, these three roots lie inside the unit circle and give three
left-decaying candidate solutions in the $B$ sector. After subtracting the two
left-boundary constraints, one left $B$-sublattice edge mode remains,
\begin{equation}
    N_L^B=3-2=1 .
\end{equation}
For $\lambda<1/2$, these roots lie outside the unit circle, so there is no left
$B$-sublattice edge mode. Therefore the root-transfer path moves the protected
left-edge zero mode from the $A$-sublattice sector to the $B$-sublattice sector
across the multicritical point, while keeping the edge-mode number unchanged along
each critical line.

\textit{Interpolation paths used in Fig.~1}---We specify the interpolation paths used
in Fig.~1. We denote by
\begin{equation}
    H_r=\sum_n\left(a_n^\dagger b_{n+r}+{\rm h.c.}\right)
\end{equation}
the range-$r$ hopping between the $A$ and $B$ sublattices.

For Fig.~1(a), we use the minimal interpolation between the $\alpha=0$ and
$\alpha'=1$ critical lines,
\begin{equation}
    H^{(a)}(\lambda)
    =
    (1-\lambda)(H_0+H_1)
    +
    \lambda (H_1+H_2),
    \qquad 0\leq \lambda\leq 1 .
\end{equation}
Thus $\lambda=0$ gives the $\alpha=0$ critical line, $\lambda=1$ gives the
$\alpha'=1$ critical line, and $\lambda=1/2$ gives the $z_{\rm dyn}=2$ Lifshitz
multicritical point.

For Fig.~1(b), we use the auxiliary-root representation of the off-diagonal
Bloch element $v_k$ to describe a transition between the $\alpha'$ and $\alpha$
critical lines, with $\alpha'=\alpha+\Delta\alpha$. Writing
$\beta\equiv e^{ik}$, we express the corresponding auxiliary function as
\begin{widetext}
\begin{equation}
    v_{\alpha,\alpha'}(\beta;\lambda)
    =
    \underbracket{\beta^\alpha}_{\alpha\ \text{roots at origin}}\times
    \underbracket{(\beta-1)}_{\text{original critical root}}\times
    \underbracket{\left[(1-\lambda)\beta-\lambda\right]^{\Delta\alpha}}
    _{\Delta\alpha\ \text{roots move from inside to outside}},
    \qquad 0\leq\lambda\leq1 .
\end{equation}
\end{widetext}
At $\lambda=0$, this gives
$v_{\alpha,\alpha'}\sim\beta^{\alpha'}(\beta-1)$, corresponding to the
$\alpha'$ critical line. At $\lambda=1$, it gives
$v_{\alpha,\alpha'}\sim\beta^\alpha(\beta-1)$ up to an overall sign,
corresponding to the $\alpha$ critical line. The moving roots are located at
$\beta=\lambda/(1-\lambda)$ and collide with the original critical root
$\beta=1$ at $\lambda=1/2$. Therefore,
\begin{equation}
    v_{\alpha,\alpha'}(\beta;1/2)
    \propto
    \beta^\alpha(\beta-1)^{\Delta\alpha+1},
\end{equation}
so the critical root becomes $(\Delta\alpha+1)$-fold degenerate and the
multicritical point has $z_{\rm dyn}=\Delta\alpha+1$.

For Fig.~1(c), the conventional Lifshitz multicritical point is represented by the
transverse-field XY-chain path
\begin{equation}
    (h(\lambda),\gamma(\lambda))
    =
    \bigl(1+\lambda\Theta(-\lambda),\,\lambda\Theta(\lambda)\bigr),
    \qquad -1\leq \lambda\leq 1 ,
\end{equation}
where $\Theta(x)$ is the Heaviside step function. Hence $\lambda=-1$ lies on the
XY critical line, $\lambda=1$ lies on the Ising critical line, and the two meet at
the Lifshitz point $(h,\gamma)=(1,0)$ at $\lambda=0$.

\let\addcontentsline\oldaddcontentsline
\onecolumngrid

\clearpage
\newpage

\widetext

\begin{center}
\textbf{\large Supplemental Material for ``Topologically Enforced Lifshitz Multicriticality in One Dimension''}
\end{center}

\maketitle

\renewcommand{\thefigure}{S\arabic{figure}}
\setcounter{figure}{0}
\renewcommand{\theequation}{S\arabic{equation}}
\setcounter{equation}{0}
\renewcommand{\thesection}{\Roman{section}}
\setcounter{section}{0}
\setcounter{secnumdepth}{4}
\counterwithout*{equation}{section}

\addtocontents{toc}{\protect\setcounter{tocdepth}{0}}
{
\tableofcontents
}

\section{$\alpha$ critical chains and their shared critical data}
\label{sm:alpha_critical_chain}

In this section, we give a brief review of the ordinary $\alpha$
critical chains. In one-dimensional chiral-symmetric free-fermion
systems, gapped SPT phases are classified by an integer winding number.
We denote a simple representative of the SPT phase with winding number
$\alpha$ by the $\alpha$ chain,
\begin{equation}
    H_\alpha
    =
    \sum_j
    \left(
        b_j^\dagger a_{j+\alpha}
        +
        {\rm h.c.}
    \right).
    \label{eq:alpha_spt_chain}
\end{equation}
The transition between two neighboring SPT phases can then be described
by the linear interpolation
\begin{equation}
    H_{\alpha,\alpha+1}(\lambda)
    =
    (1-\lambda)H_\alpha
    +
    \lambda H_{\alpha+1},
    \qquad
    0\leq \lambda\leq 1 .
    \label{eq:alpha_spt_interpolation}
\end{equation}
At $\lambda=0$ and $\lambda=1$, the Hamiltonian describes the
$\alpha$ and $\alpha+1$ SPT chains, respectively. At the symmetric point
$\lambda=1/2$, it becomes
\begin{equation}
    H_{\alpha,\alpha+1}
    \left(\lambda=\frac{1}{2}\right)
    =
    \frac{1}{2}
    \sum_j
    \left[
        b_j^\dagger a_{j+\alpha}
        +
        b_j^\dagger a_{j+\alpha+1}
        +
        {\rm h.c.}
    \right].
    \label{eq:ordinary_alpha_critical_real_space}
\end{equation}
Up to an overall normalization, this is the $\alpha$ critical chain we introduced in the main text. Thus the $\alpha$ critical chain is defined as the critical point between the neighboring $\alpha$ and $\alpha+1$ SPT phases.

\begin{figure}[t]
  \centering
  \begin{minipage}[b]{0.32\linewidth}
    \panel{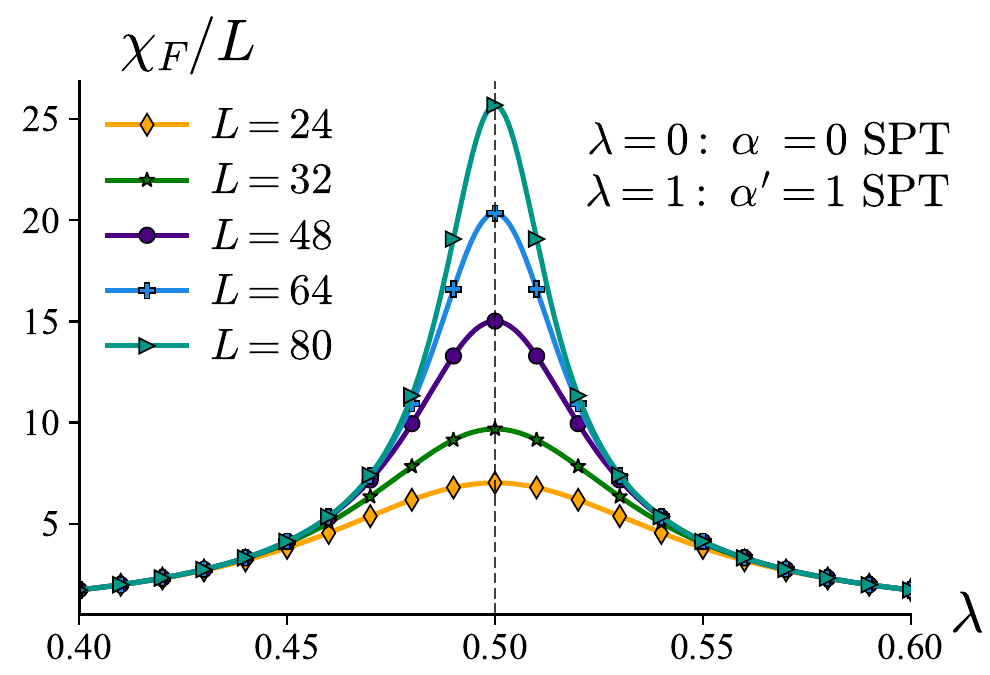}{\linewidth}{(a)}
  \end{minipage}
  \begin{minipage}[b]{0.32\linewidth}
    \panel{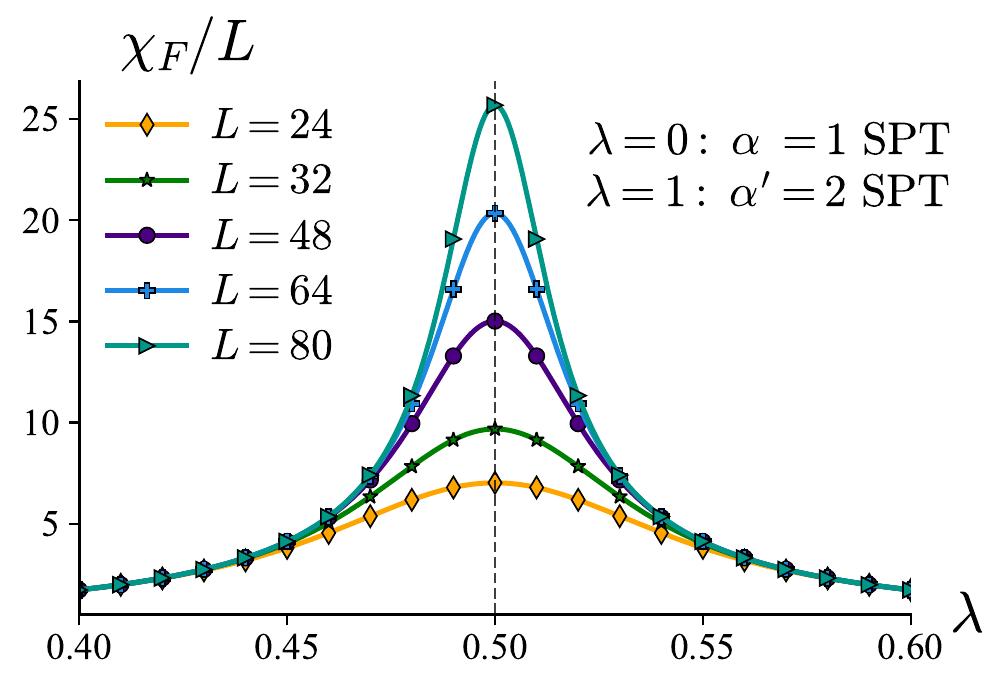}{\linewidth}{(b)}
  \end{minipage}
  \begin{minipage}[b]{0.32\linewidth}
    \panel{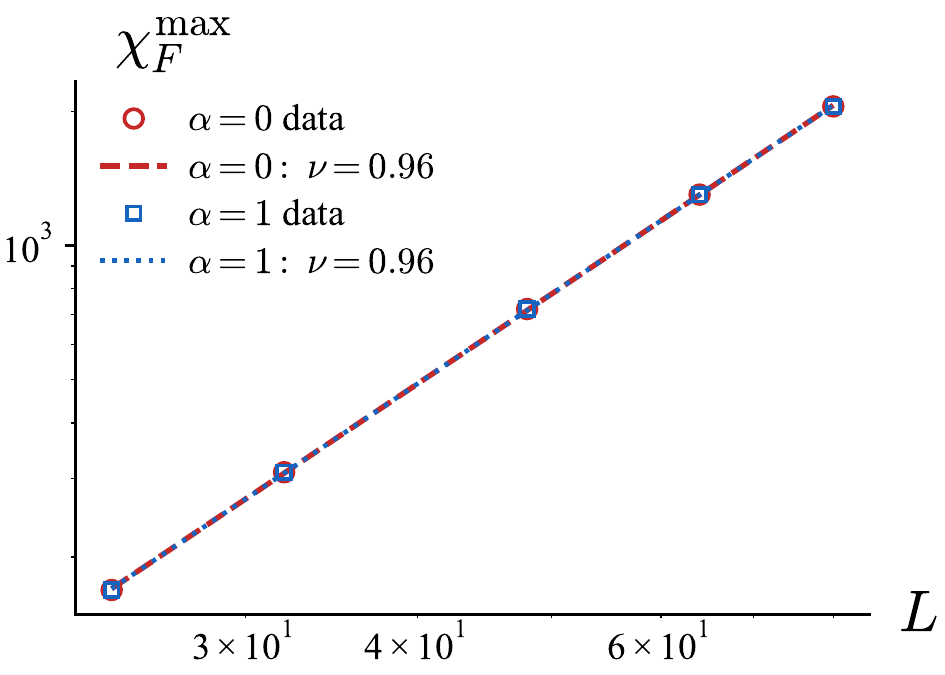}{\linewidth}{(c)}
  \end{minipage}
\caption{
Fidelity susceptibility and correlation-length exponent of ordinary
$\alpha$ critical chains.
(a) Fidelity susceptibility density $\chi_F/L$ along the linear
interpolation between the $\alpha=0$ and $\alpha=1$ SPT chains. The
dashed line marks $\lambda=1/2$, where the system realizes the
$\alpha=0$ critical chain.
(b) Fidelity susceptibility density $\chi_F/L$ along the linear
interpolation between the $\alpha=1$ and $\alpha=2$ SPT chains. The
peak again sharpens at $\lambda=1/2$, giving the $\alpha=1$ critical
chain.
(c) Finite-size scaling of the fidelity-susceptibility peak for the
$\alpha=0$ and $\alpha=1$ critical chains. The two fits give consistent
values of $\nu\simeq 1$. Together with the central-charge result shown
in the main text, this indicates that different $\alpha$ critical chains share the same
local critical data and are distinguished only by their topology.
}
\label{fig:alpha_chain_nu}
\end{figure}

We first locate this critical point using the ground-state fidelity
susceptibility along the interpolation parameter $\lambda$. For two normalized many-body ground states at nearby parameters,
$|\Psi_0(\lambda)\rangle$ and $|\Psi_0(\lambda+\delta\lambda)\rangle$,
the ground-state fidelity is defined as
\begin{equation}
F(\lambda,\lambda+\delta\lambda)
=
\left|
\left\langle
\Psi_0(\lambda)
\middle|
\Psi_0(\lambda+\delta\lambda)
\right\rangle
\right|.
\end{equation}
The corresponding fidelity susceptibility is
\begin{equation}
    \chi_F(\lambda)
    =
    -\lim_{\delta\lambda\rightarrow 0}
    \frac{
    2\ln F(\lambda,\lambda+\delta\lambda)
    }
    {(\delta\lambda)^2}.
    \label{eq:alpha_chain_fidelity_definition}
\end{equation}
As shown in Fig.~\ref{fig:alpha_chain_nu}(a,b), the fidelity
susceptibility density $\chi_F/L$ develops a sharpening peak at
$\lambda=1/2$ for both the $\alpha=0$ and $\alpha=1$ critical chains.
This confirms that the ordinary $\alpha$ critical chain occurs at the
transition point between two neighboring SPT chains. The finite-size
scaling of the peak height is then used to extract the
correlation-length exponent. For the peak of fidelity susceptibility, we have
\begin{equation}
    \chi_F^{\max}
    \sim
    L^{2/\nu},
    \label{eq:alpha_chain_fidelity_scaling_total}
\end{equation}
The comparison in Fig.~\ref{fig:alpha_chain_nu}(c) shows that the
$\alpha=0$ and $\alpha=1$ critical chains give consistent values of
$\nu\simeq 1$.

We now describe the same critical chain in momentum space. At
$\lambda=1/2$, the single-particle Bloch Hamiltonian takes the
off-diagonal form
\begin{equation}
    \mathcal H_\alpha(k)
    =
    \begin{pmatrix}
        0 & v_\alpha^*(k) \\
        v_\alpha(k) & 0
    \end{pmatrix},
    \label{eq:ordinary_alpha_critical_bloch}
\end{equation}
with
\begin{equation}
    v_\alpha(k)
    =
    e^{i\alpha k}
    \left(1+e^{ik}\right).
    \label{eq:ordinary_alpha_critical_vk}
\end{equation}
The spectrum is therefore
\begin{equation}
    E_\alpha(k)
    =
    \pm |v_\alpha(k)|.
\end{equation}
Expanding around the gap-closing momentum $k=\pi+q$, one obtains
\begin{equation}
    1+e^{ik}
    =
    1-e^{iq}
    \simeq
    -iq \quad\Longrightarrow\quad E_\alpha(q)
    \sim
    |q|.
\end{equation}
Thus all ordinary $\alpha$ critical chains have
\begin{equation}
    z=1 .
\end{equation}
Together with the fidelity result $\nu\simeq 1$, this shows that these
critical chains share the same conventional critical exponents. The
central charge has already been extracted from the entanglement-entropy
scaling in the main text, where the ordinary $\alpha$ critical chains
were found to have
\begin{equation}
    c=1 .
\end{equation}
The only difference between these critical chains is their topology. We now
write
\begin{equation}
    \beta=e^{ik},
\end{equation}
so that the off-diagonal element at the ordinary $\alpha$ critical point
becomes
\begin{equation}
    v_\alpha(\beta)
    =
    \beta^\alpha(\beta+1).
    \label{eq:ordinary_alpha_critical_beta}
\end{equation}
In this form, the factor $\beta^\alpha$ encodes the SPT index inherited
from the neighboring gapped phases. As reviewed in the End Matter, the
OBC boundary-mode content can be determined from the root structure of
$v(\beta)$, so the $\alpha$ critical chain host $\min\{|\alpha|,|\alpha+1|\}$ pairs of topological edge modes under OBC. The generalized Li--Haldane correspondence\cite{guo2025generalizedlihaldanecorrespondencecritical} then relates the
same topological information to the number of corresponding midgap
structures in the periodic-boundary entanglement spectrum.

Therefore, the ordinary $\alpha$ critical chains share the same local
critical data,
\begin{equation}
    z=1,
    \qquad
    \nu =  1,
    \qquad
    c=1,
\end{equation}
and are distinguished only by their topology. The phase transition between different
$\alpha$ critical chains discussed in this work is therefore not a
conventional transition between distinct critical universality classes.
Rather, it is enforced by the topological mismatch between critical
chains that otherwise have the same local critical behavior.

\section{Lattice Hamiltonians for topologically enforced multicritical points}
\label{sm:mcp_lattice_construction}

In this section, we construct
lattice Hamiltonians that realize multicritical points between
topologically distinct critical chains. Using the $\beta$ notation
introduced in the SM Sec.~\ref{sm:alpha_critical_chain}, the $\alpha$ critical chain
is represented by
\begin{equation}
    v_\alpha(\beta)
    =
    \beta^\alpha(\beta+1).
\end{equation}
We consider two critical chains labeled by $\alpha$ and
$\alpha'=\alpha+\Delta\alpha$, with $\Delta\alpha>0$:
\begin{equation}
    v_\alpha(\beta)
    =
    \beta^\alpha(\beta+1),
    \qquad
    v_{\alpha'}(\beta)
    =
    \beta^{\alpha+\Delta\alpha}(\beta+1).
\end{equation}
As discussed above, these critical chains have the same local critical
data, but differ in their topological root structure. In particular,
they share the same critical root at $\beta=-1$, while the
$\alpha'$ critical chain contains $\Delta\alpha$ additional roots at
$\beta=0$ compared with the $\alpha$ critical chain.

Changing from the $\alpha'$ critical chain to the $\alpha$ critical
chain therefore requires transferring these $\Delta\alpha$ topological
roots across the unit circle. Along a generic interpolation, the moving
roots need not cross the unit circle at the same momentum or at the
same tuning parameter. Such a path can produce several separate gap
closings and hence several low-energy sectors. In this work, we focus
instead on the case where all topology-changing roots collide with the
pre-existing critical root at $\beta=-1$ at a single tuning parameter.
This produces a single higher-order Lifshitz multicritical point, as
illustrated schematically in
Fig.1(b) in main text.

A simple interpolation that realizes this root collision is
\begin{equation}
    v_{\alpha,\alpha'}(\beta;\lambda)
    =
    \beta^\alpha(\beta+1)
    \big[(1-\lambda)\beta+\lambda\big]^{\Delta\alpha},
    \qquad
    0\leq \lambda\leq 1 .
    \label{eq:mcp_interpolate}
\end{equation}
At $\lambda=0$, this reduces to
\begin{equation}
    v_{\alpha,\alpha'}(\beta;0)
    =
    \beta^{\alpha+\Delta\alpha}(\beta+1)
    =
    v_{\alpha'}(\beta),
\end{equation}
while at $\lambda=1$ it becomes
\begin{equation}
    v_{\alpha,\alpha'}(\beta;1)
    =
    \beta^\alpha(\beta+1)
    =
    v_\alpha(\beta).
\end{equation}
The moving roots are located at
\begin{equation}
    \beta_\lambda
    =
    -\frac{\lambda}{1-\lambda}.
\end{equation}
They collide with the original critical root $\beta=-1$ at
$\lambda=1/2$. Therefore, the multicritical point is described by
\begin{equation}
    v^{\rm mc}_{\alpha,\alpha'}(\beta)
    \propto
    \beta^\alpha(\beta+1)^{\Delta\alpha+1}.
    \label{eq:mcp_vk}
\end{equation}

\begin{figure}[t]
  \centering
  \begin{minipage}[b]{0.36\linewidth}
    \panel{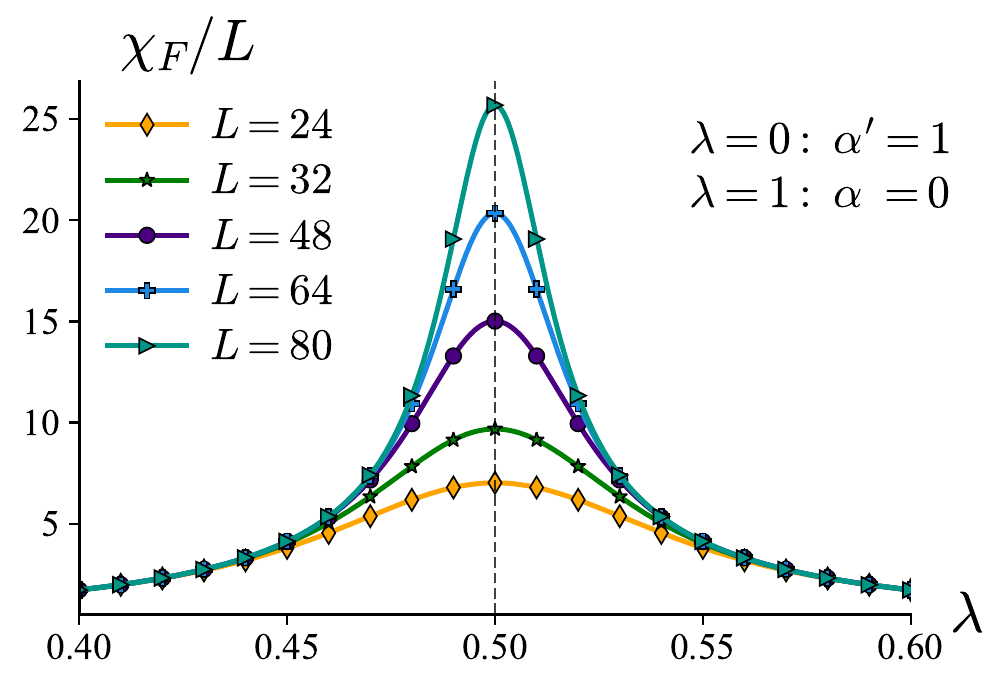}{\linewidth}{(a)}
  \end{minipage}
  \begin{minipage}[b]{0.36\linewidth}
    \panel{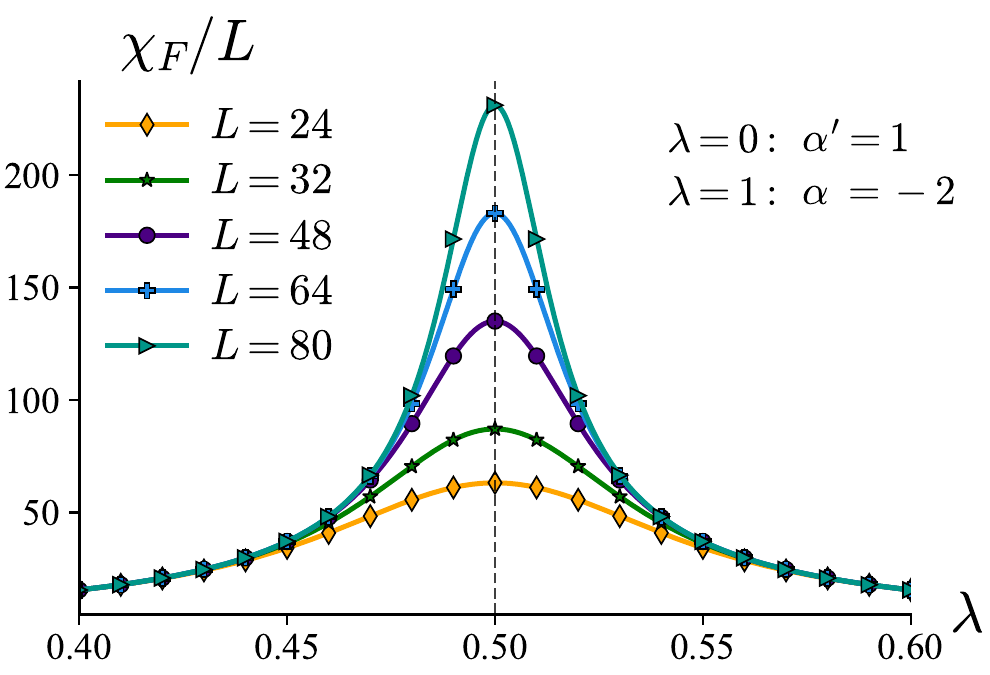}{\linewidth}{(b)}
  \end{minipage}
\caption{
Fidelity susceptibility for topologically enforced Lifshitz multicritical points. 
(a) Fidelity susceptibility density $\chi_F/L$ as a function of the
interpolation parameter $\lambda$ for the minimal case
$\alpha=0$ and $\alpha'=1$. The dashed line marks the multicritical
point $\lambda_c=1/2$, where the moving root
$\beta_\lambda=-\lambda/(1-\lambda)$ collides with the fixed critical
root at $\beta=-1$. The sharpening peak identifies the topology-changing
multicritical point. 
(b) Fidelity susceptibility density $\chi_F/L$ for the interpolation between $\alpha=-2$ and $\alpha'=1$. In this case, $\Delta\alpha=3$, so three moving roots collide with the fixed critical root at $\lambda_c=1/2$. The peak at the same value of $\lambda$ demonstrates that the root-collision construction continues to identify the Lifshitz multicritical point even when the interpolation connects negative- and positive-$\alpha$ critical lines.
}
\label{fig:mcp_interpolate}
\end{figure}

Expanding around the collision momentum $k=\pi+q$, one has
\begin{equation}
    \beta+1
    =
    e^{ik}+1
    =
    1-e^{iq}
    \simeq
    -iq .
\end{equation}
Hence the dispersion near the multicritical point behaves as
\begin{equation}
    E(q)
    \sim
    |v^{\rm mc}_{\alpha,\alpha'}(q)|
    \sim
    |q|^{\Delta\alpha+1},
\end{equation}
so that
\begin{equation}
    z_{\rm dyn}
    =
    \Delta\alpha+1 .
\end{equation}
The dynamical exponent is therefore fixed by the order of the root
collision.

We further characterize the same multicritical point through the
fidelity susceptibility along the interpolation parameter $\lambda$,
using the same definition as in the previous section. As shown in
Fig.~\ref{fig:mcp_interpolate}(b), $\chi_F/L$ develops a pronounced
peak at $\lambda_c=1/2$, where the moving root collides with the fixed
critical root at $\beta=-1$. The peak becomes sharper with increasing
system size, confirming that the topology-changing point is located at
the root-collision point.

In momentum space, the full interpolation is obtained by setting
$\beta=e^{ik}$:
\begin{equation}
v_{\alpha,\alpha'}(k;\lambda)
=
e^{i\alpha k}(1+e^{ik})
\big[(1-\lambda)e^{ik}+\lambda\big]^{\Delta\alpha}.
\end{equation}
Using the binomial expansion,
\begin{equation}
\big[(1-\lambda)e^{ik}+\lambda\big]^{\Delta\alpha}
=
\sum_{s=0}^{\Delta\alpha}
\frac{\Delta\alpha!}{s!(\Delta\alpha-s)!}
(1-\lambda)^s\lambda^{\Delta\alpha-s}e^{isk}, 
\end{equation}
we obtain
\begin{equation}
v_{\alpha,\alpha'}(k;\lambda)
=
\sum_{s=0}^{\Delta\alpha}
\frac{\Delta\alpha!}{s!(\Delta\alpha-s)!}
(1-\lambda)^s\lambda^{\Delta\alpha-s}
\left[
e^{i(\alpha+s)k}
+
e^{i(\alpha+s+1)k}
\right].
\end{equation}
Therefore the corresponding real-space Hamiltonian is
\begin{equation}
H_{\alpha,\alpha'}(\lambda)
=
\sum_j
\sum_{s=0}^{\Delta\alpha}
\frac{\Delta\alpha!}{s!(\Delta\alpha-s)!}
(1-\lambda)^s\lambda^{\Delta\alpha-s}
\left[
a_j^\dagger b_{j+\alpha+s}
+
a_j^\dagger b_{j+\alpha+s+1}
+
{\rm h.c.}
\right].
\end{equation}
At the multicritical point $\lambda=1/2$, this reduces to
\begin{equation}
v^{\rm mc}_{\alpha,\alpha'}(k)
\propto
e^{i\alpha k}(1+e^{ik})^{\Delta\alpha+1}.
\end{equation}
Equivalently,
\begin{equation}
H^{\rm mc}_{\alpha,\alpha'}
=
\sum_j
\sum_{r=0}^{\Delta\alpha+1}
\frac{(\Delta\alpha+1)!}{r!(\Delta\alpha+1-r)!}
\left(
a_j^\dagger b_{j+\alpha+r}
+
{\rm h.c.}
\right),
\end{equation}
up to an overall normalization.
Thus the binomial hopping structure is not imposed by hand; it follows
directly from requiring all topology-changing roots to collide with the
original critical root at a single multicritical point.

We emphasize that this construction is different from a generic linear
interpolation between the two critical-chain Hamiltonians. In the
minimal case $\alpha=0$ and $\alpha'=1$, the root-collision construction
can be realized by a simple linear interpolation between the two
critical lines. For general $\alpha$ and $\alpha'$, however, a naive
linear interpolation does not in general force all moving roots to
collide with the original critical root at the same parameter. The
root-based construction above is therefore the natural way to define
the desired single Lifshitz multicritical point.

The root-based construction also has a broader range of validity than the most naive root-counting picture. In particular, it can naturally treat transitions between critical lines with negative and positive $\alpha$. For a fixed negative $\alpha$, the root structure can be described equivalently using the reverse variable $\tilde\beta=\beta^{-1}$. However, when the transition connects a negative-$\alpha$ critical line to a positive-$\alpha'$ critical line, the intuitive picture of simply moving ordinary roots from inside to outside the $\beta$ unit circle is no longer sufficient. Nevertheless, the interpolation Eq.~\eqref{eq:mcp_interpolate} is still well defined as a Laurent-polynomial construction and continues to realize the desired root collision at $\lambda=1/2$.

As an illustrative example, we consider the transition between
$\alpha=-2$ and $\alpha'=1$, for which $\Delta\alpha=3$. The
interpolation is
\begin{equation}
v_{-2,1}(\beta;\lambda)
=
\beta^{-2}(\beta+1)
\big[(1-\lambda)\beta+\lambda\big]^3 .
\end{equation}
At $\lambda=0$, one obtains
\begin{equation}
v_{-2,1}(\beta;0)
=
\beta(\beta+1)
=
v_1(\beta),
\end{equation}
whereas at $\lambda=1$,
\begin{equation}
v_{-2,1}(\beta;1)
=
\beta^{-2}(\beta+1)
=
v_{-2}(\beta).
\end{equation}
The three moving roots are located at
\begin{equation}
\beta_\lambda
=
-\frac{\lambda}{1-\lambda},
\end{equation}
and collide with the original critical root $\beta=-1$ at
$\lambda=1/2$. Therefore this path still produces a single Lifshitz
multicritical point, now with
\begin{equation}
z_{\rm dyn}
=
\Delta\alpha+1
=
4 .
\end{equation}
We verify this construction numerically by computing the
ground-state fidelity susceptibility along the same interpolation path.
As shown in Fig.~\ref{fig:mcp_interpolate}(c), the fidelity
susceptibility develops a peak at $\lambda=1/2$, precisely where the
three moving roots collide with the original critical root. This
provides a direct numerical check that the same root-collision
construction identifies the Lifshitz multicritical point even for a
transition between negative and positive $\alpha$ critical lines.

\section{Absence of topological edge modes in one dimensional conventional multicritical points}
\label{sm2}


As a topologically trivial benchmark, we consider the transverse-field XY
chain
\begin{equation}
H_{\mathrm{t}\rm XY}
=
-\frac12\sum_j
\left[
(1+\gamma)X_jX_{j+1}
+
(1-\gamma)Y_jY_{j+1}
+
2h Z_j
\right].
\end{equation}
Using the Jordan--Wigner transformation,
\begin{equation}
Z_j=1-2c_j^\dagger c_j,
\qquad
X_j+iY_j
=
2\left(\prod_{\ell<j} Z_\ell\right)c_j ,
\end{equation}
and neglecting the boundary-parity term in the thermodynamic limit, the
Hamiltonian becomes a quadratic fermion Hamiltonian. In momentum space it
can be written in the BdG form
\begin{equation}
H
=
\frac12\sum_k
\Psi_k^\dagger
\mathcal H(k)
\Psi_k,
\qquad
\Psi_k=
\begin{pmatrix}
c_k\\
c^\dagger_{-k}
\end{pmatrix},
\end{equation}
with
\begin{equation}
\mathcal H(k)
=
\xi_k\tau_z+\Delta_k\tau_y,
\qquad
\xi_k=h-\cos k,
\qquad
\Delta_k=\gamma\sin k .
\end{equation}
The corresponding excitation energy is
\begin{equation}
E_k
=
2\epsilon_k,
\qquad
\epsilon_k
=
\sqrt{\xi_k^2+\Delta_k^2}
=
\sqrt{(h-\cos k)^2+\gamma^2\sin^2 k}.
\end{equation}
The phase diagram contains two critical lines. The Ising critical line is
obtained at
\begin{equation}
h=1,\qquad \gamma\neq0,
\end{equation}
and has central charge \(c=1/2\). The XY critical line is obtained at
\begin{equation}
\gamma=0,\qquad |h|<1,
\end{equation}
and has central charge \(c=1\). These two critical lines meet at
\begin{equation}
(h,\gamma)=(1,0).
\end{equation}
At this point,
\begin{equation}
E_k=2(1-\cos k)\sim k^2 ,
\end{equation}
so the point is a conventional \(z=2\) Lifshitz multicritical point.

At the conventional Lifshitz point \((h,\gamma)=(1,0)\), the pairing term
vanishes and the BdG Hamiltonian reduces to
\begin{equation}
H_{\rm Lif}
=
\sum_k
(1-\cos k)
\left(
c_k^\dagger c_k
-
c_{-k}c_{-k}^\dagger
\right),
\end{equation}
up to an overall normalization convention. 

Fourier transforming to real space gives
\begin{equation}
H_{\rm Lif}
=
\sum_j
\left[
c_j^\dagger c_j
-
c_j c_j^\dagger
\right]
-
\frac12
\sum_j
\left[
c_j^\dagger c_{j+1}
+
c_{j+1}^\dagger c_j
-
c_j c_{j+1}^\dagger
-
c_{j+1}c_j^\dagger
\right].
\end{equation}
Equivalently, using the fermion anticommutation relation, this can be
written as
\begin{equation}
H_{\rm Lif}
=
\sum_j
\left[
(2c_j^\dagger c_j-1)
-
\left(
c_j^\dagger c_{j+1}
+
c_{j+1}^\dagger c_j
\right)
\right].
\end{equation}
The first term is not an irrelevant detail here: it is the onsite
particle-hole contribution of the BdG Hamiltonian and will become the
onsite Majorana coupling.

We now introduce Majorana operators
\begin{equation}
a_j=c_j+c_j^\dagger,
\qquad
b_j=-i(c_j-c_j^\dagger),
\end{equation}
so that
\begin{equation}
c_j=\frac12(a_j+i b_j),
\qquad
c_j^\dagger=\frac12(a_j-i b_j).
\end{equation}
The onsite BdG term becomes
\begin{equation}
2c_j^\dagger c_j-1
=
i a_j b_j .
\end{equation}
The nearest-neighbor hopping term becomes
\begin{equation}
c_j^\dagger c_{j+1}
+
c_{j+1}^\dagger c_j
=
\frac{i}{2}
\left(
a_j b_{j+1}
+
a_{j+1}b_j
\right).
\end{equation}
Therefore the Lifshitz-point Hamiltonian takes the Majorana form
\begin{equation}
H_{\rm Lif}
=
i\sum_j
\left[
a_j b_j
-
\frac12 a_j b_{j+1}
-
\frac12 a_{j+1}b_j
\right].
\end{equation}
Equivalently, after relabeling the last term,
\begin{equation}
H_{\rm Lif}
=
\frac{i}{2}\sum_j
a_j
\left[
2b_j
- b_{j+1}
- b_{j-1}
\right].
\end{equation}
Thus the corresponding Majorana coupling function up to a propotional factor is
\begin{equation}
v_{\rm Lif}(k)
=
2-e^{ik}-e^{-ik}
=
2(1-\cos k)
\sim
k^2.
\label{eq:conven_vk}
\end{equation}
Therefore the conventional Lifshitz point realizes a symmetric
second-difference coupling in the Majorana representation.  This is the coupling structure as the topologically trivial \(z=2\) Lifshitz
theory: the coupling extends symmetrically to the left and right and does not have a nontrivial winding.

\section{Robustness against symmetry-preserving disorder and interactions}
\label{sm3}

\subsection{Robustness against symmetry-preserving disorder}

We now examine whether the topologically enforced Lifshitz multicritical point
and its associated entanglement signatures remain stable against disorder. To
introduce disorder in a symmetry-preserving way, it is useful to first rewrite
the single-particle Hamiltonian in the explicitly chiral basis
\begin{equation}
    (A_0,A_1,\cdots,A_{L-1},B_0,B_1,\cdots,B_{L-1}) ,
\end{equation}
rather than the real-space interleaved basis
\begin{equation}
    (A_0,B_0,A_1,B_1,\cdots).
\end{equation}
This is only a reordering of basis states, but it makes the off-diagonal
chiral structure manifest and avoids introducing disorder terms that
accidentally break the chiral symmetry. In this basis, the Hamiltonian takes the
form
\begin{equation}
    H_\lambda
    =
    \begin{pmatrix}
        0 & D_\lambda^\dagger \\
        D_\lambda & 0
    \end{pmatrix}.
\end{equation}
The interpolation is encoded entirely in the off-diagonal block. For the minimal case we considered here, it is given by
\begin{equation}
    D_\lambda^{(0)}
    =
    \left[(1-\lambda)I-\lambda T\right](I-T),
\end{equation}
where \(T\) is the lattice shift operator defined by \(T|j\rangle=|j+1\rangle\).
In matrix form,
\begin{equation}
    T
    =
    \begin{pmatrix}
        0 & 0 & \cdots & \eta_{\rm bc} \\
        1 & 0 & \cdots & 0 \\
        0 & 1 & \ddots & \vdots \\
        \vdots & \ddots & \ddots & 0
    \end{pmatrix},
\end{equation}
where \(\eta_{\rm bc}=1\) for PBC, \(\eta_{\rm bc}=-1\) for APBC, and
\(\eta_{\rm bc}=0\) for OBC.

For PBC, this interpolation has a simple momentum-space interpretation. At
\(\lambda=0\), one obtains
\begin{equation}
    D_{\lambda=0}^{(0)}=I-T,
    \qquad
    v_k=1-e^{ik}.
\end{equation}
At \(\lambda=1\), one obtains
\begin{equation}
    D_{\lambda=1}^{(0)}=-T(I-T),
    \qquad
    v_k=-e^{ik}(1-e^{ik}).
\end{equation}
Therefore, the interpolation connects $\alpha=0$ critical chain to $\alpha' = 1$ critical chain. At
the midpoint \(\lambda=1/2\), the two factors become identical up to an overall
constant, giving the clean \(z=2\) Lifshitz operator,
\begin{equation}
    D_{\lambda=1/2}^{(0)}
    =
    \frac{1}{2}(I-T)^2 .
\end{equation}
The disorder is then introduced as a random stiffness in the factorized
off-diagonal block,
\begin{equation}
    D_\lambda
    =
    \left[(1-\lambda)I-\lambda T\right]J(I-T),
    \qquad
    J=\mathrm{diag}(J_1,\cdots,J_L),
\end{equation}
with
\begin{equation}
    J_j=1+\delta_j,
    \qquad
    \delta_j\in[-W,W].
\end{equation}
We take \(W<1\), so that all \(J_j\) remain positive. This disorder breaks
translation symmetry, but preserves the chiral off-diagonal structure and the
random-stiffness form of the Lifshitz operator.

\begin{figure}[t]
  \centering
  \begin{minipage}[b]{0.245\linewidth}
    \panel{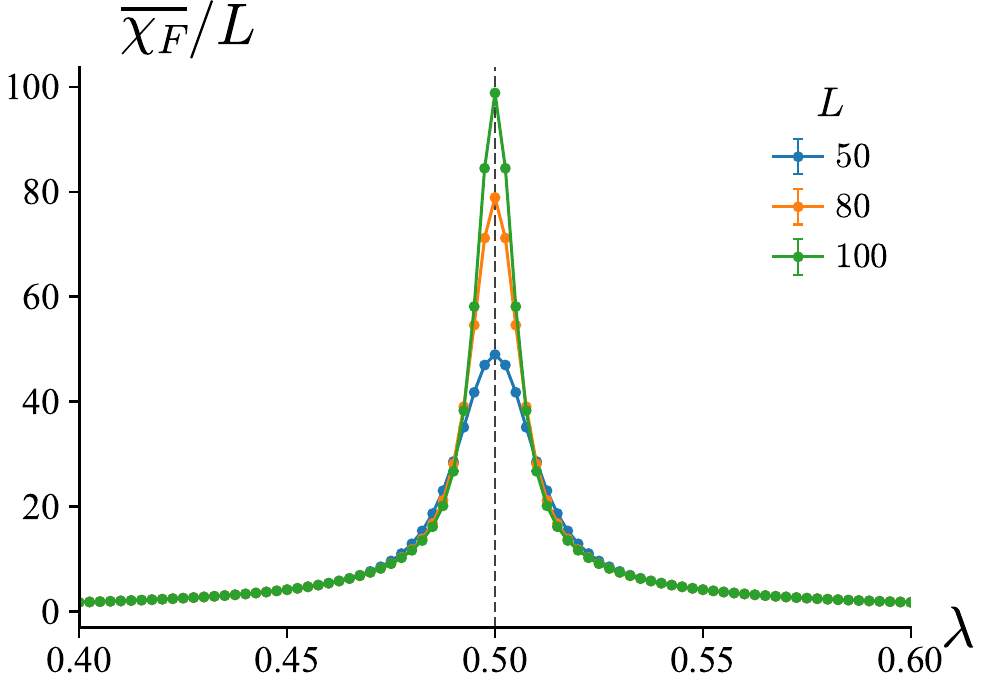}{\linewidth}{(a)}
  \end{minipage}
  \begin{minipage}[b]{0.245\linewidth}
    \panel{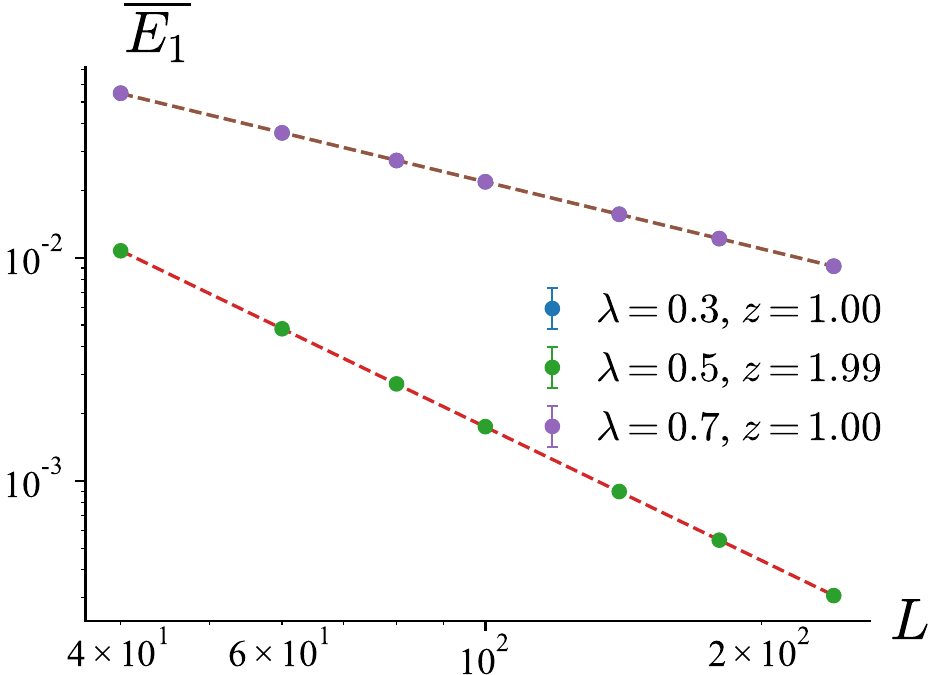}{\linewidth}{(b)}
  \end{minipage}
  \begin{minipage}[b]{0.245\linewidth}
    \panel{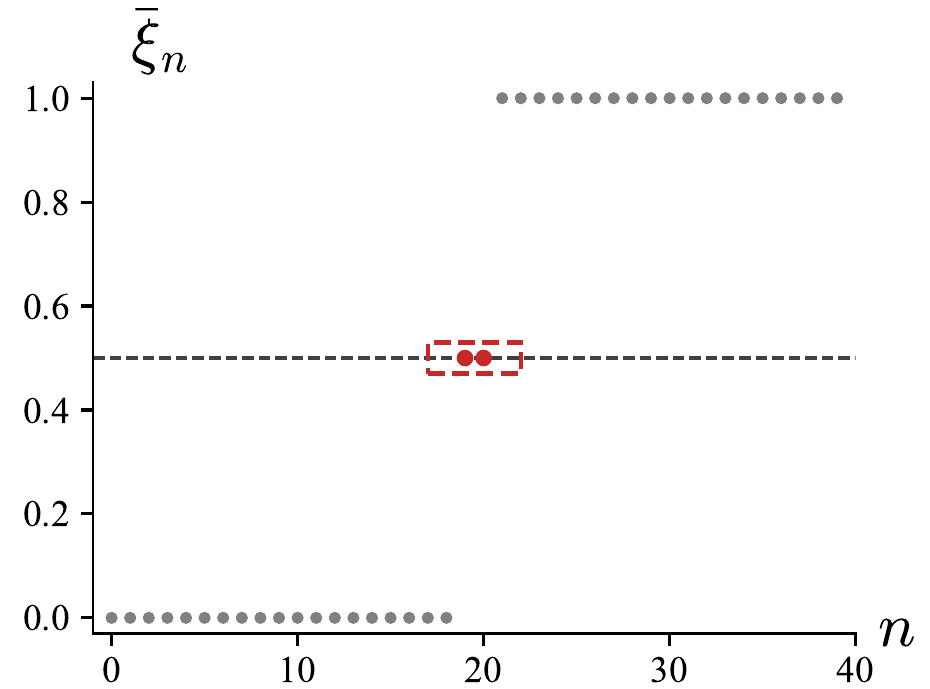}{\linewidth}{(c)}
  \end{minipage}
  \begin{minipage}[b]{0.245\linewidth}
    \panel{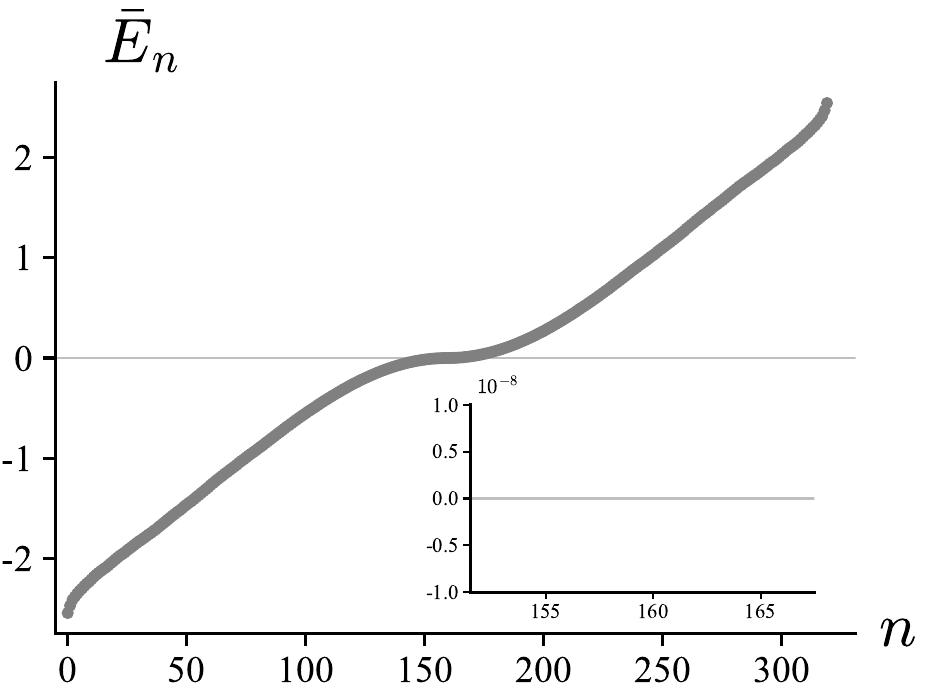}{\linewidth}{(d)}
  \end{minipage}
\caption{
Robustness of the topologically enforced Lifshitz multicritical point under
random-stiffness disorder.
(a) Disorder-averaged fidelity susceptibility
\(\overline{\chi_F}/L\) along the interpolation
\(D_\lambda=[(1-\lambda)I-\lambda T]J(I-T)\). The peak remains centered at
\(\lambda=1/2\), showing that the random-stiffness disorder does not shift the
multicritical point.
(b) Disorder-averaged first nonzero PBC energy \(\overline{E_1}\) as a function
of system size. The fitted exponent in \(\overline{E_1}\sim L^{-z}\) is close
to \(z\simeq1\) for \(\lambda=0.3\) and \(\lambda=0.7\), while it approaches
\(z\simeq2\) at \(\lambda=0.5\), confirming the Lifshitz nature of the
disordered MCP.
(c) Disorder-averaged PBC entanglement spectrum at the disordered Lifshitz MCP
for \(L=100\) and \(L_A=20\). The two red midgap modes remain pinned near
\(\bar{\xi}_n=1/2\), showing that the topology-related entanglement signature
is robust against disorder.
(d) Disorder-averaged full signed OBC single-particle spectrum at the same
disordered MCP. The inset shows a \(10^{-8}\) zero-mode window around
\(\bar{E}_n=0\), which remains empty, indicating the absence of protected
physical OBC zero modes. Thus the PBC entanglement midgap modes are not
accompanied by physical OBC zero modes, and the Li--Haldane mismatch persists
under disorder.
The disorder is generated by \(J_j=1+\delta_j\), with
\(\delta_j\in[-W,W]\) and \(W=0.5\). All data are averaged over
\(N_{\rm dis}=100\) disorder realizations. Error bars denote the standard error
of the mean; in panels (c) and (d), they are smaller than the marker size and
are therefore not visible.
}
\label{fig:disorder}
\end{figure}

Before testing the topological entanglement signature, we first verify that this
disorder does not shift the location of the Lifshitz multicritical point. In
the clean limit, the interpolation connects two adjacent critical chains. Away
from the midpoint, only one factor becomes gapless at long wavelength, giving a
\(z=1\) critical theory. At
\begin{equation}
    \lambda=\frac12,
\end{equation}
the two factors become identical up to an overall constant,
\begin{equation}
    D_{\lambda=1/2}
    =
    \frac12 (I-T)J(I-T),
\end{equation}
which is the random-stiffness version of the \(z=2\) Lifshitz operator. Since
\(J\) is positive and invertible, it changes the local stiffness profile but
does not remove the coalescence of the two difference operators. Thus, within
this disorder class, the multicritical point is expected to remain fixed at
\(\lambda=1/2\).

This expectation is confirmed numerically by the disorder-averaged fidelity
susceptibility shown in Fig.~\ref{fig:disorder}(a). For each disorder
realization, the same stiffness profile \(J\) is used for all values of
\(\lambda\), so that the fidelity compares ground states within the same
disordered sample. The averaged susceptibility \(\overline{\chi_F}/L\) develops
a sharp peak centered at \(\lambda=1/2\), and the peak becomes more pronounced
with increasing system size. This indicates that the random-stiffness disorder
does not shift the multicritical point.

We further confirm the Lifshitz nature of this disordered multicritical point
from the finite-size scaling of the first nonzero PBC energy,
\begin{equation}
    E_1(L)\sim L^{-z}.
\end{equation}
As shown in Fig.~\ref{fig:disorder}(b), the fitted exponent is close to
\(z\simeq1\) away from the midpoint, while at \(\lambda=1/2\) it approaches
\(z\simeq2\). Therefore, both the fidelity susceptibility and the dynamical
scaling identify the same unshifted Lifshitz multicritical point.

Having established that the multicritical point remains at \(\lambda=1/2\), we
now test the robustness of the topology-related entanglement signature directly
at this disordered Lifshitz point. Up to an unimportant overall factor, the
relevant operator is
\begin{equation}
    D_W=(I-T)J(I-T).
\end{equation}
Under PBC, the entanglement spectrum remains characterized by midgap levels at
\(\xi=1/2\). As shown in Fig.~\ref{fig:disorder}(c), for \(L=100\) and
subsystem size \(L_A=20\), the two midgap entanglement modes remain pinned near
\(\xi=1/2\) after disorder averaging. This shows that the topology-related
entanglement signature is not an artifact of translation symmetry.

We then compare this PBC entanglement signature with the physical OBC spectrum.
If the usual Li--Haldane correspondence held in the naive form, the midgap
structure in the PBC entanglement spectrum would be accompanied by protected
physical boundary zero modes under OBC. However, the full signed OBC spectrum in
Fig.~\ref{fig:disorder}(d) shows no such exponentially small boundary modes. In
particular, the inset displays a \(10^{-8}\) zero-mode window around
\(E=0\), which remains empty. The low-energy OBC states are therefore not
protected boundary zero modes, but ordinary finite-size Lifshitz modes.

These results establish the stability of the Li--Haldane mismatch under the
symmetry-preserving disorder considered here. First, the random-stiffness
disorder does not shift the topologically enforced Lifshitz multicritical point
away from \(\lambda=1/2\). Second, once evaluated at this unshifted disordered
MCP, the PBC entanglement spectrum retains its topology-related midgap
structure, while the physical OBC spectrum still lacks protected exponentially
small edge modes. Hence the breakdown of the naive Li--Haldane correspondence
is robust against moderate random-stiffness disorder.

\subsection{Robustness against symmetry-preserving interaction}

We further test whether the Li--Haldane mismatch at the topologically
enforced Lifshitz MCP remains visible after adding interactions. We focus on the minimal case \(\alpha=0\) and \(\alpha'=1\), namely
\(\Delta\alpha=1\), of the topologically enforced Lifshitz MCP introduced in
Eq.~\eqref{eq:mcp_interpolate}. The free part can be written in the factorized form
\begin{equation}
H_0(\lambda)
=
\sum_j
\left[
\mathcal K_j(\lambda)+\mathcal K_j^\dagger(\lambda)
\right],
\end{equation}
with
\begin{equation}
\mathcal K_j(\lambda)
=
\left[(1-\lambda)b_j^\dagger-\lambda b_{j-1}^\dagger\right]
(a_j-a_{j+1}) .
\end{equation}
After shifting the dummy summation index, one obtains
\begin{equation}
\sum_j \mathcal K_j(\lambda)
=
\sum_j b_j^\dagger
\left[
(1-\lambda)a_j-a_{j+1}+\lambda a_{j+2}
\right].
\end{equation}
At \(\lambda=1/2\), this reduces to
\begin{equation}
\sum_j \mathcal K_j(1/2)
=
\frac{1}{2}
\sum_j
b_j^\dagger
\left(
a_j-2a_{j+1}+a_{j+2}
\right),
\end{equation}
which realizes the \(z_{\rm dyn}=2\) Lifshitz MCP, up to the harmless
staggered-gauge choice that fixes whether the double root is placed at
\(k=0\) or \(k=\pi\).

We then introduce the interacting deformation
\begin{equation}
H(\lambda,U)
=
\sum_j
\left[
\mathcal K_j(\lambda)+\mathcal K_j^\dagger(\lambda)
\right]
+
U\sum_j
:\mathcal K_j^\dagger(\lambda)\mathcal K_j(\lambda): .
\end{equation}
Here \(:\cdots:\) denotes fermionic normal ordering. Namely, after expanding
\(\mathcal K_j^\dagger \mathcal K_j\), all creation operators are moved to the
left of all annihilation operators, and the contraction terms generated by the
fermion anticommutation relations are subtracted. Therefore
\(:\mathcal K_j^\dagger \mathcal K_j:\) keeps only the genuine quartic
interaction part, rather than also adding extra quadratic hopping terms. The
interaction is thus constructed from the same local factor \(\mathcal K_j\)
that generates the Lifshitz MCP. In the numerical calculation below, we take
\(U=0.5\).

For the DMRG calculation, we order the two sublattice fermions as a single
one-dimensional chain,
\begin{equation}
c_{2j}=a_j,\qquad c_{2j+1}=b_j .
\end{equation}
Equivalently, this fermionic chain can be represented by a spin-\(1/2\) chain
through the Jordan--Wigner transformation
\begin{equation}
c_m
=
\left(\prod_{\ell<m} Z_\ell\right)\sigma_m^-,
\qquad
c_m^\dagger
=
\left(\prod_{\ell<m} Z_\ell\right)\sigma_m^+,
\qquad
n_m=c_m^\dagger c_m .
\end{equation}
The resulting model is a finite-range \(U(1)\)-conserving interacting chain,
which we implement as a matrix-product operator. We use this interacting DMRG
setup to compare the many-body PBC entanglement spectrum with the physical
OBC low-energy spectrum.

\begin{figure}[t] 
\centering 
\begin{minipage}[b]{0.245\linewidth} 
\panel{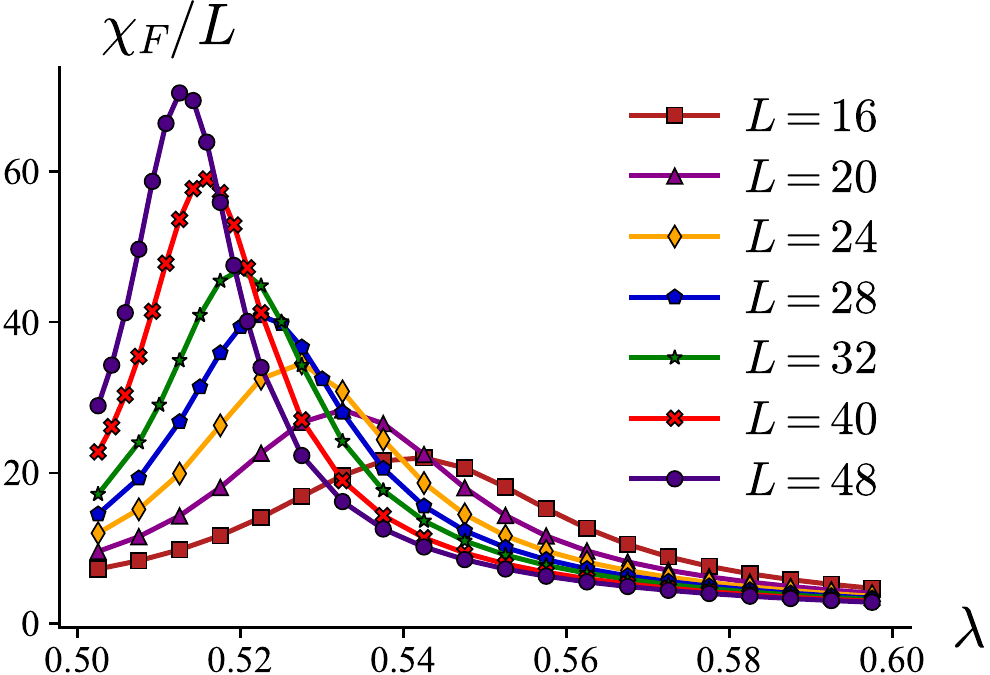}{\linewidth}{(a)} 
\end{minipage} 
\begin{minipage}[b]{0.245\linewidth} 
\panel{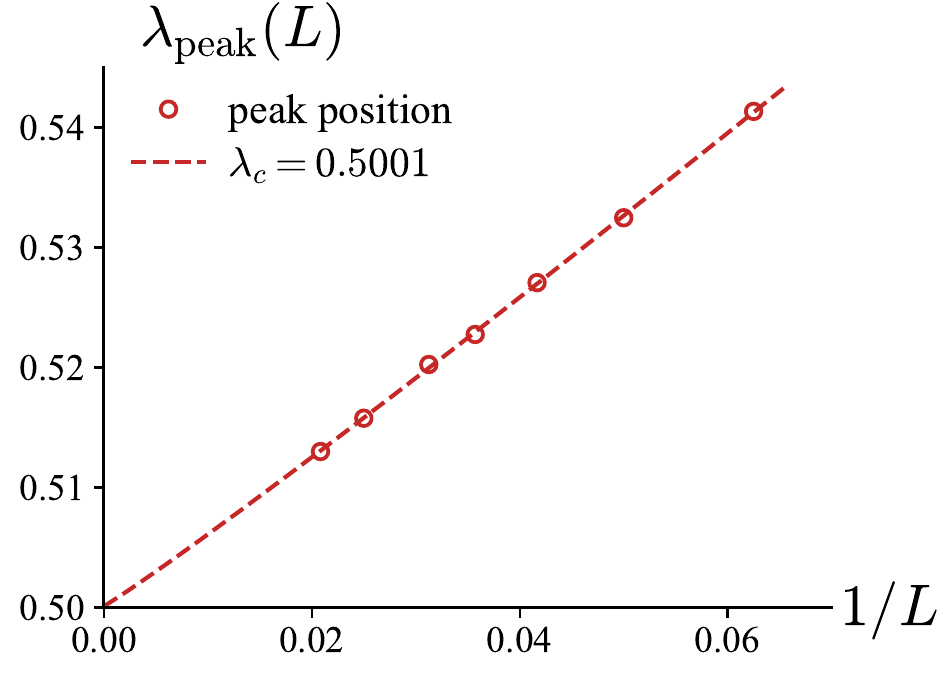}{\linewidth}{(b)} 
\end{minipage} 
\begin{minipage}[b]{0.245\linewidth} 
\panel{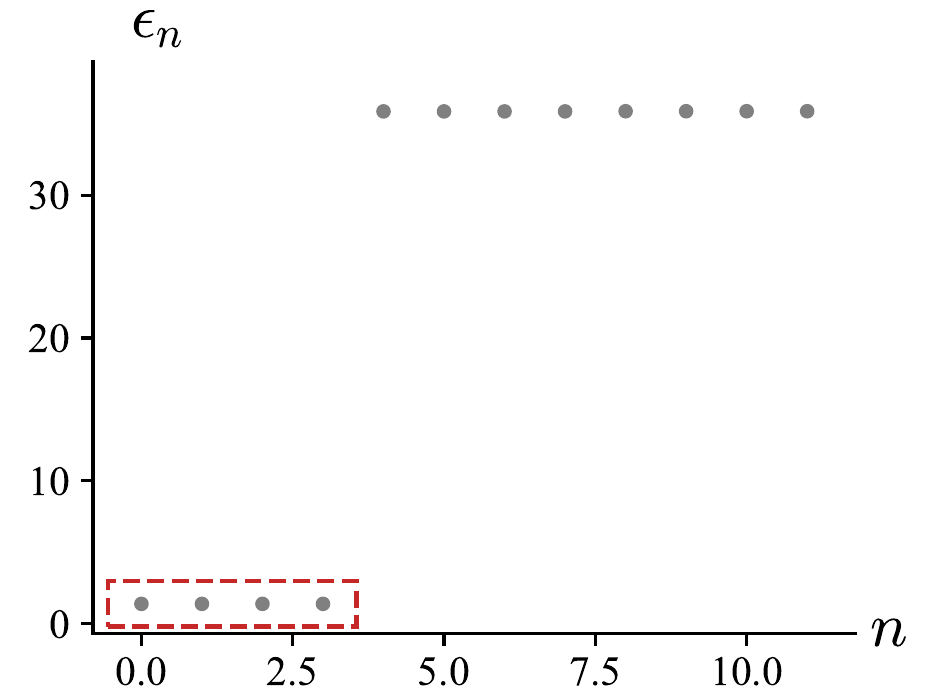}{\linewidth}{(c)} 
\end{minipage} 
\begin{minipage}[b]{0.245\linewidth} 
\panel{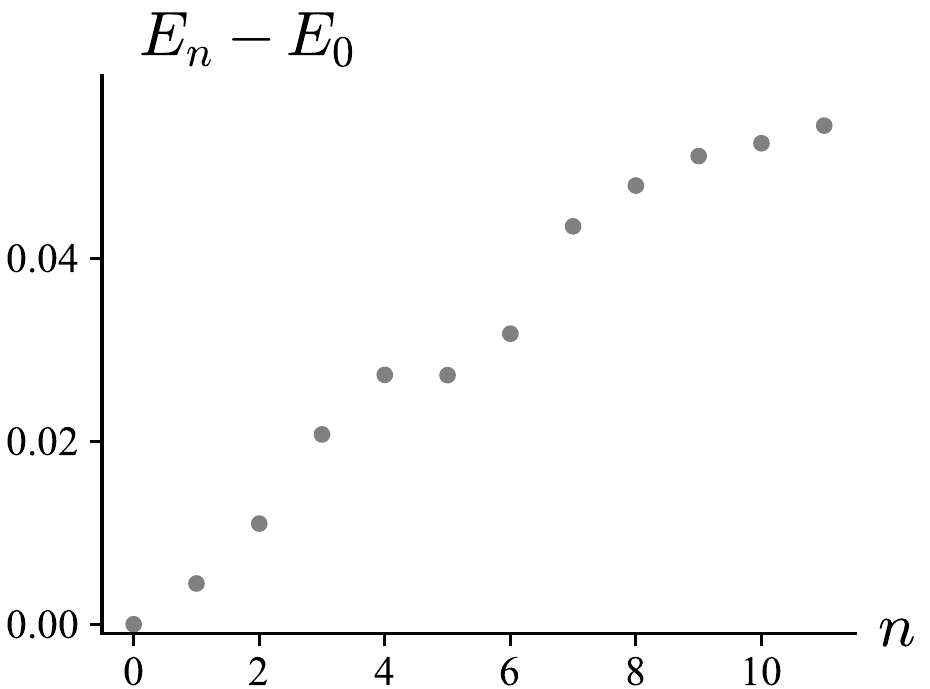}{\linewidth}{(d)} 
\end{minipage} 
\caption{ Interacting diagnostic of the Li--Haldane mismatch at the topologically enforced Lifshitz multicritical point. (a) Fidelity susceptibility density $\chi_F/L$ as a function of the interpolation parameter $\lambda$ for different system sizes at $U=0.5$. The peak sharpens near $\lambda=1/2$ as $L$ increases, indicating the interacting topology-changing point. (b) Finite-size extrapolation of the peak position $\lambda_{\rm peak}(L)$. The hollow red circles denote the finite-size peak positions extracted from the fidelity susceptibility, while the red dashed line shows a fit to $\lambda_{\rm peak}(L)=\lambda_c+aL^{-b}$. The extrapolated value of $\lambda_c$ is close to the expected multicritical point $\lambda=1/2$. (c) Many-body PBC entanglement spectrum at the interacting MCP. The four lowest entanglement levels, highlighted by the red dashed box, form a nearly degenerate low-lying structure with $\epsilon_n\simeq\log 4$. (d) Low-energy many-body spectrum under OBC at the same parameters. No corresponding protected zero-energy boundary multiplet is observed. Thus the interacting system retains the same qualitative Li--Haldane mismatch as the free-fermion topologically enforced MCP: the PBC entanglement spectrum exhibits a robust topology-related degeneracy, whereas the physical OBC spectrum does not show the matching boundary degeneracy. } \label{fig:interaction} \end{figure}

We first verify that the interacting topology-changing point remains
pinned near $\lambda=1/2$. Using the fidelity susceptibility defined
above, we compute $\chi_F(\lambda,L)$ along the interacting interpolation
at $U=0.5$. As shown in Fig.~\ref{fig:interaction}(a), $\chi_F/L$
develops a pronounced peak near $\lambda=1/2$, and the peak becomes
sharper as the system size is increased. We use the peak only as a
diagnostic of the transition location, and do not extract a
correlation-length exponent from the peak-height scaling in the
interacting case.

To estimate the thermodynamic-limit transition point, we track the
finite-size peak position $\lambda_{\rm peak}(L)$ and fit its drift by
the empirical form
\begin{equation}
    \lambda_{\rm peak}(L)
    =
    \lambda_c
    +
    a L^{-b}.
\end{equation}
As shown in Fig.~\ref{fig:interaction}(b), the extrapolated value of
$\lambda_c$ is very close to $\lambda=1/2$. This supports that the
interacting system retains the topology-enforced Lifshitz multicritical
point near the same root-collision point as in the free-fermion limit.

Having identified the interacting MCP, we next compare the bulk entanglement
spectrum under PBC with the physical low-energy spectrum under OBC. Since the
system is interacting, the free-fermion correlation-matrix spectrum is no
longer available. We therefore directly compute the many-body Schmidt spectrum
of the DMRG ground state. For the PBC calculation, the physical Hamiltonian is
periodic and the system is bipartitioned into two equal halves. The
entanglement energies are defined by
\begin{equation}
\epsilon_n=-\log \rho_n ,
\end{equation}
where \(\rho_n\) are the Schmidt weights.

As shown in Fig.~\ref{fig:interaction}(d), the PBC entanglement spectrum
exhibits a clear fourfold low-lying structure. This should be understood as
the many-body counterpart of the two midgap entanglement modes found in the
free-fermion correlation spectrum in the main text. In the free-fermion
description, each midgap correlation level \(\xi=1/2\) corresponds to an
entanglement zero mode with two possible occupations. Therefore, two such
midgap modes generate a \(2^2=4\)-fold degeneracy in the many-body Schmidt
spectrum. Consistently, in the interacting calculation the four leading
Schmidt weights are nearly equal,
\begin{equation}
\rho_0\simeq \rho_1\simeq \rho_2\simeq \rho_3\simeq \frac{1}{4},
\end{equation}
or equivalently
\begin{equation}
\epsilon_0\simeq\epsilon_1\simeq\epsilon_2\simeq\epsilon_3\simeq \log 4 .
\end{equation}
The corresponding entanglement entropy is therefore close to
\begin{equation}
S_A\simeq \log 4 .
\end{equation}
Thus, although the correlation-matrix diagnostic is no longer available in
the interacting system, the same topology-related entanglement structure
survives as a fourfold degeneracy in the many-body PBC entanglement spectrum.

We then compare this bulk entanglement feature with the physical OBC spectrum.
If the usual Li--Haldane correspondence applied in its naive form, the
fourfold PBC entanglement structure would be accompanied by a corresponding
protected low-energy boundary multiplet under OBC. However, as shown in
Fig.~\ref{fig:interaction}(e), the OBC many-body spectrum does not exhibit such a
protected zero-energy multiplet. Instead, the excited states remain separated
from the ground state by finite-size gaps. Thus the interacting calculation
shows the same qualitative mismatch as the free-fermion result: the PBC
entanglement spectrum retains a robust degeneracy, whereas the physical OBC
spectrum does not contain the corresponding protected boundary degeneracy.

These results indicate that the Li--Haldane breakdown at the topologically
enforced Lifshitz MCP is not merely an artifact of the free-fermion
correlation-matrix description. The interaction preserves the essential
separation between the virtual entanglement-boundary structure and the
physical open-boundary spectrum. Together with the disorder analysis, this
supports the robustness of the Li--Haldane mismatch at the topologically
enforced Lifshitz multicritical point.

\section{Detailed discussion of the breakdown of the Li–Haldane correspondence}
\label{sm4}




\subsection{Numerical details for the $\alpha=0$ family with general $\alpha'$}

In the main text, we used the minimal case $\alpha=0$ and $\alpha'=1$ to demonstrate the breakdown of the Li--Haldane correspondence at a topologically enforced Lifshitz multicritical point. Here we provide numerical details for the $\alpha=0$ family with general $\alpha'$. As derived in Eq.~\eqref{eq:mcp_vk}, the multicritical Bloch element reduces to
\begin{equation}
v^{\rm mc}_{0,\alpha'}(\beta)
\sim
(\beta-1)^z,
\qquad
z=\alpha'+1,
\qquad
\beta=e^{ik}.
\label{eq:mc_beta_alpha0}
\end{equation}
Thus the Lifshitz exponent is fixed by the topological separation from the $\alpha=0$ critical line. Since Eq.~\eqref{eq:mc_beta_alpha0} contains no additional root at $\beta=0$, this family provides a clean setting for testing whether the Lifshitz root itself can generate physical OBC zero modes.

\begin{figure}[t]
  \centering
  \begin{minipage}[b]{0.32\linewidth}
    \panel{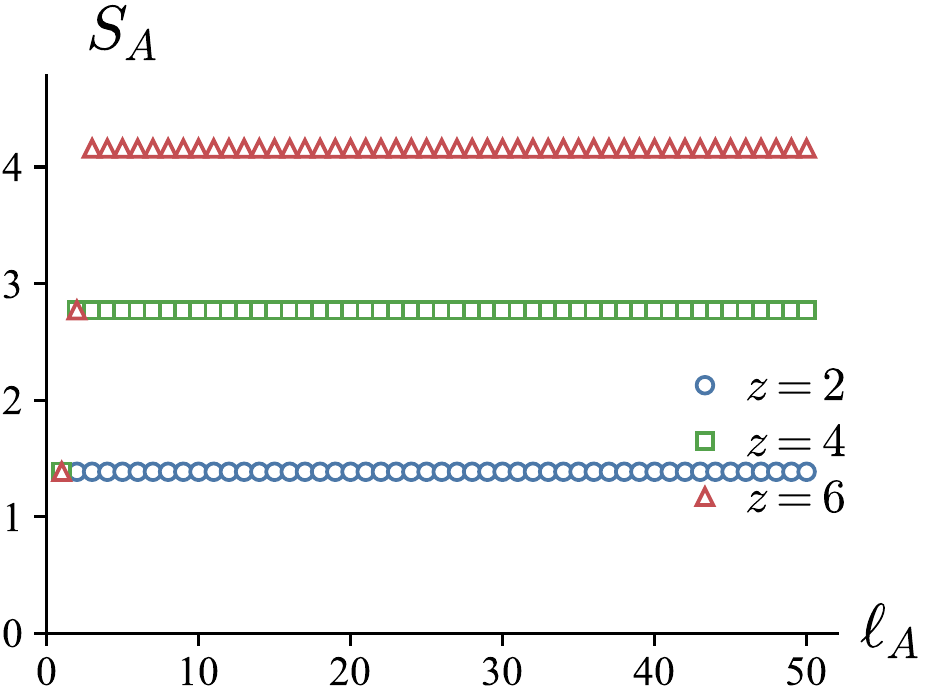}{\linewidth}{(a)}
  \end{minipage}
  \begin{minipage}[b]{0.32\linewidth}
    \panel{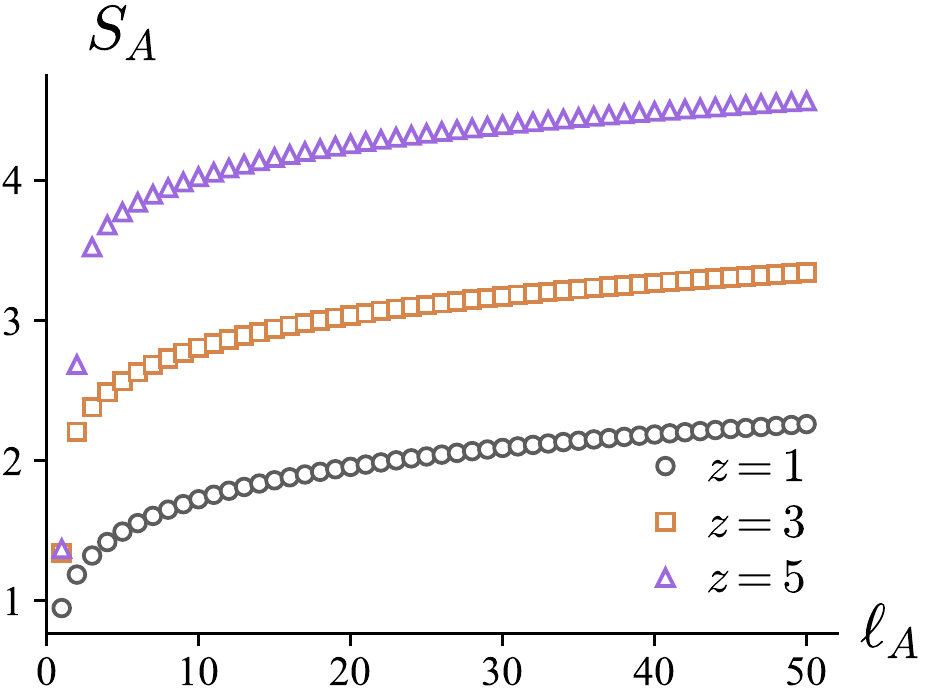}{\linewidth}{(b)}
  \end{minipage}
  \begin{minipage}[b]{0.32\linewidth}
    \panel{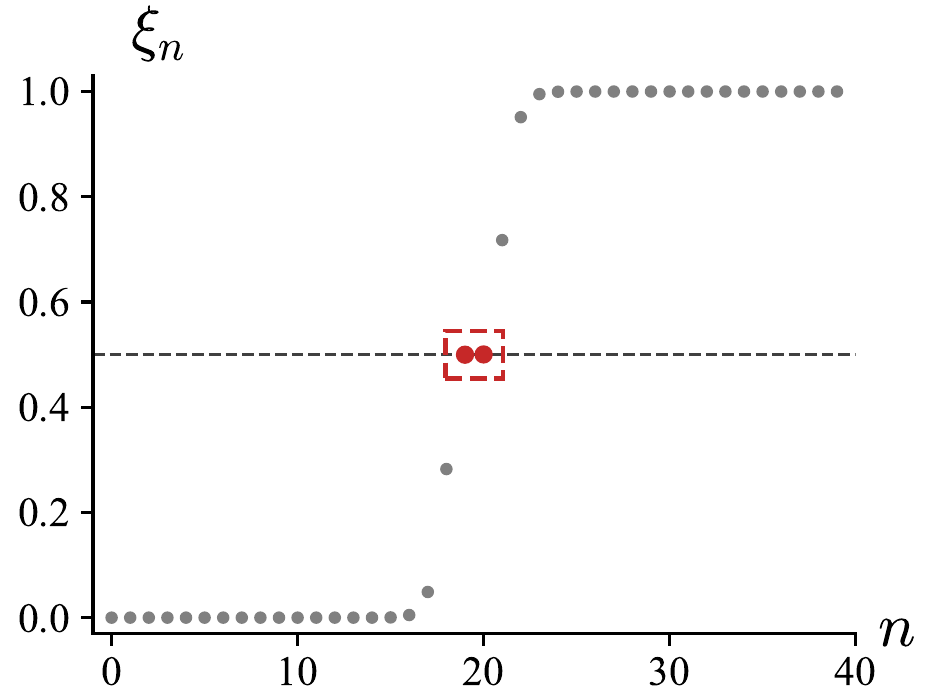}{\linewidth}{(c)}
  \end{minipage}
  \begin{minipage}[b]{0.32\linewidth}
    \panel{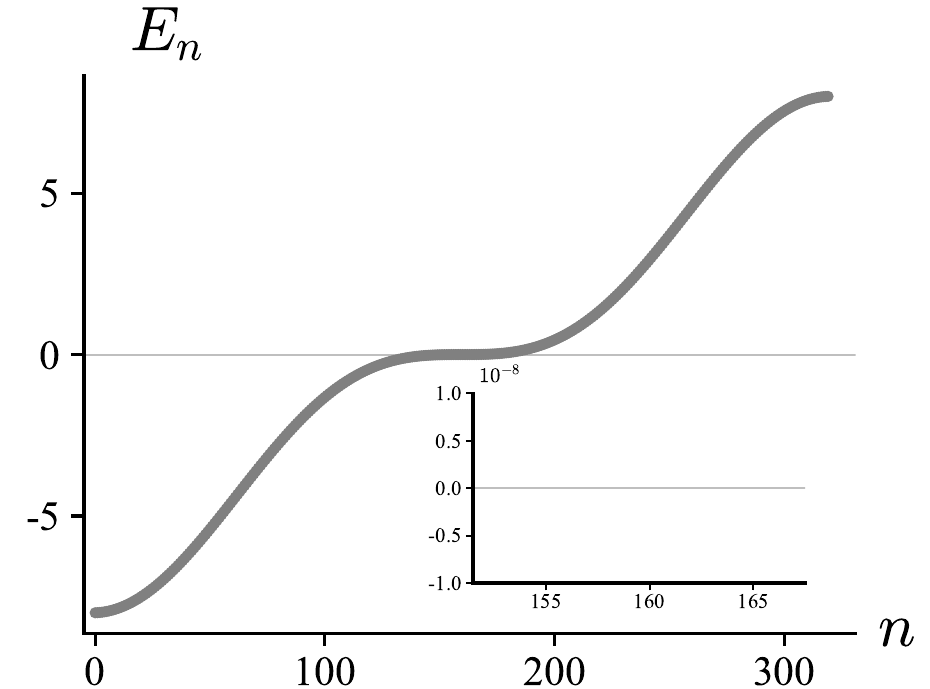}{\linewidth}{(d)}
  \end{minipage}
  \begin{minipage}[b]{0.32\linewidth}
    \panel{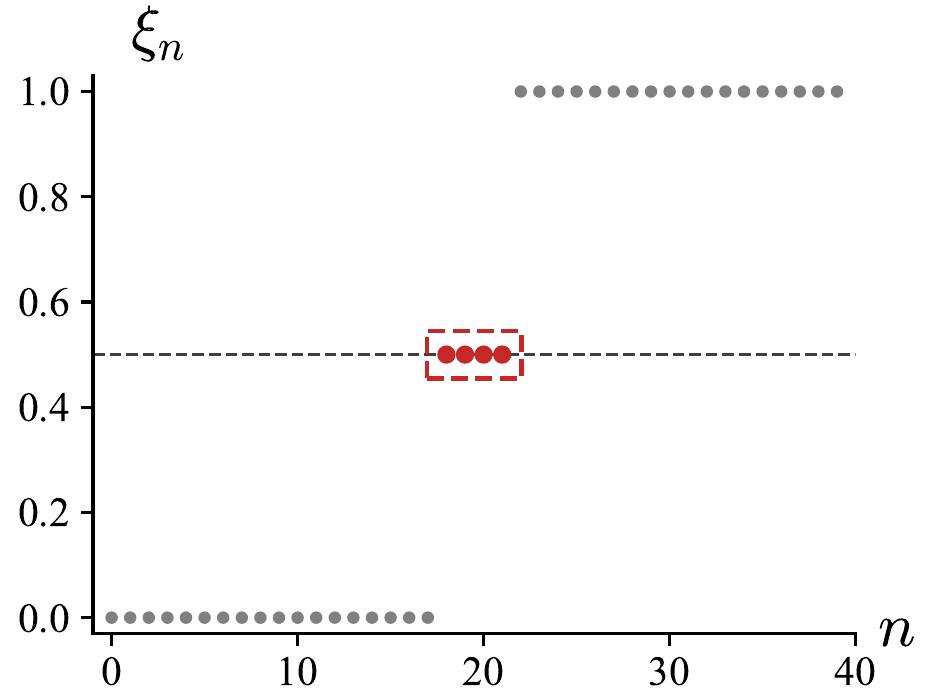}{\linewidth}{(e)}
  \end{minipage}
  \begin{minipage}[b]{0.32\linewidth}
    \panel{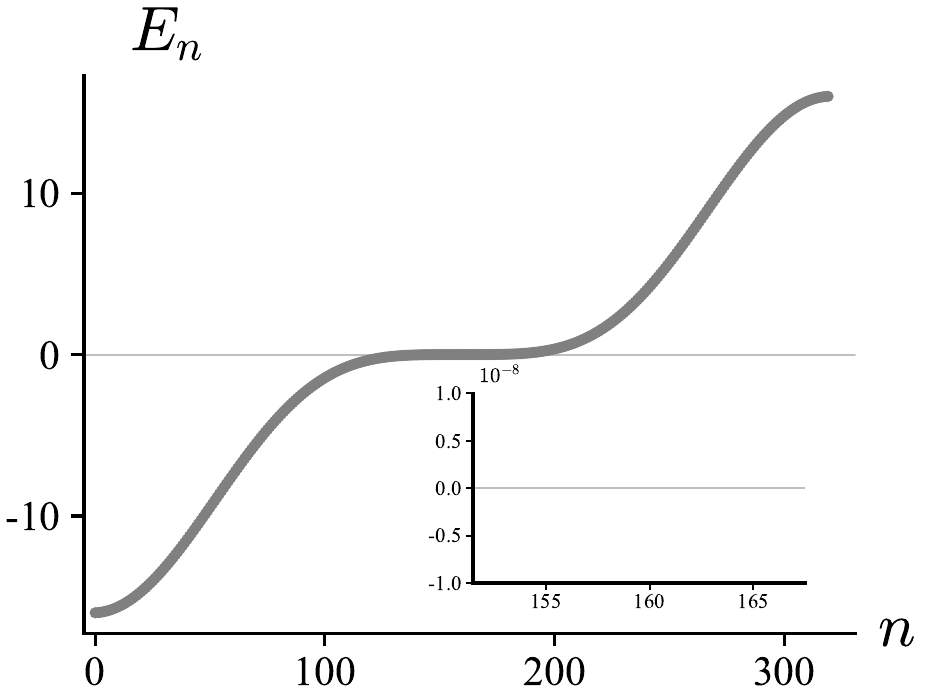}{\linewidth}{(f)}
  \end{minipage}
\caption{
Numerical evidence for the Li--Haldane mismatch in the $\alpha=0$ Lifshitz family.
(a) Entanglement entropy $S_A$ for even Lifshitz exponents $z=2,4,6$. The entropy rapidly saturates with increasing subsystem size $\ell_A$, and the plateau values are consistent with $S_A\simeq z\ln 2$, indicating $z/2$ midgap pairs in the PBC correlation spectrum.
(b) Entanglement entropy for odd Lifshitz exponents $z=1,3,5$. In contrast to the even-$z$ cases, $S_A$ shows long-range logarithmic growth, similar to the ordinary $z=1$ critical chain.
(c,d) Representative odd Lifshitz point with $z=3$. The PBC correlation spectrum contains one midgap pair pinned near $\xi_n=1/2$, highlighted by the red dashed box in (c), while the physical OBC spectrum in (d) shows no isolated exponentially small zero-energy boundary mode. The inset magnifies the $10^{-8}$ energy window around zero.
(e,f) Representative even Lifshitz point with $z=4$. The PBC correlation spectrum contains two midgap pairs near $\xi_n=1/2$, consistent with the plateau in (a), whereas the OBC spectrum again contains no isolated zero mode in the $10^{-8}$ window.
These results show that the Lifshitz root can generate midgap entanglement levels without producing corresponding physical OBC zero modes, demonstrating the breakdown of the Li--Haldane correspondence at topologically enforced Lifshitz multicriticality.
}
\label{fig:numeric}
\end{figure}

The main numerical distinction is between even and odd $z$. For even $z$, the entanglement entropy displays a short-range, gapped-like behavior, although the physical spectrum remains gapless at the Lifshitz point. As shown in Fig.~\ref{fig:numeric}(a), the curves for $z=2,4,6$ rapidly saturate as the subsystem size is increased. The saturated values are quantized in units of $\ln 2$ and are consistent with
\begin{equation}
S_A \simeq z\ln 2 .
\end{equation}
This quantized plateau suggests the presence of $z$ individual midgap levels, or equivalently $z/2$ midgap pairs, in the PBC correlation spectrum.

For odd $z$, the behavior is qualitatively different. As shown in Fig.~\ref{fig:numeric}(b), the entanglement entropy for $z=1,3,5$ exhibits long-range logarithmic scaling similar to the ordinary $z=1$ critical chain. This indicates that odd-$z$ Lifshitz points retain a long-range correlation structure similar to that of the $z=1$ critical point.

We next compare the PBC entanglement spectrum with the physical OBC spectrum. For even Lifshitz MCPs within the $\alpha=0$ family, the numerical pattern is summarized by
\begin{equation}
N^{\rm ES}_{\rm pair} = \frac{z}{2},
\qquad
N^{\rm OBC}_{\rm pair} = 0 ,
\qquad
z\in 2\mathbb{Z}.
\label{eq:pattern_even}
\end{equation}
The minimal $z=2$ case has already been shown in the main text. As a higher-order example, Fig.~\ref{fig:numeric}(e) shows that the $z=4$ PBC correlation spectrum contains two midgap pairs pinned at $\xi=1/2$, consistent with the entropy plateau in Fig.~\ref{fig:numeric}(a). However, the corresponding OBC spectrum in Fig.~\ref{fig:numeric}(f) does not contain isolated exponentially small zero-energy modes.

For odd Lifshitz MCPs, the numerical pattern is instead
\begin{equation}
N^{\rm ES}_{\rm pair} = 
\frac{z-1}{2},
\qquad
N^{\rm OBC}_{\rm pair} = 
0 ,
\qquad
z\in 2\mathbb{Z}+1 .
\label{eq:pattern_odd}
\end{equation}
The $z=3$ example is shown in Figs.~\ref{fig:numeric}(c) and \ref{fig:numeric}(d). The PBC correlation spectrum contains one midgap pair at $\xi=1/2$, whereas the OBC spectrum shows no isolated zero-energy boundary mode.

These numerical results indicate that the breakdown of the Li--Haldane correspondence is a general feature of topologically enforced Lifshitz multicritical points within the $\alpha=0$ family. In the following subsections, we provide an analytic explanation for this numerical observation and show how the mismatch follows from the distinction between the flattened entanglement kernel and the unflattened OBC boundary-root problem.

\subsection{Even-$z$ Lifshitz MCPs: winding and entanglement midgap modes}

We now explain analytically the even-$z$ numerical results for the $\alpha=0$ family. The multicritical Bloch element is
\begin{equation}
v_z(k)\sim (e^{ik}-1)^z .
\end{equation}
We first show that the midgap modes in the entanglement spectrum are not accidental, but are topological through the winding number. Recall the definition of the winding number,
\begin{equation}
W
=
\frac{1}{2\pi i}
\int_{\rm BZ} dk\,
v_k^{-1}\partial_k v_k .
\end{equation}
To see which part of $v_k$ contributes to this quantity, we decompose the Bloch element into its amplitude and phase,
\begin{equation}
v_k=|v_k|q_k,
\qquad
q_k=\frac{v_k}{|v_k|}.
\end{equation}
Then the winding can be rewritten as
\begin{equation}
W
=
\frac{1}{2\pi i}
\int_{\rm BZ} dk\,
\left(
\partial_k\log |v_k|
+
q_k^{-1}\partial_k q_k
\right)=
\frac{1}{2\pi i}
\int_{\rm BZ} dk\,
q_k^{-1}\partial_k q_k ,
\label{eq:winding_q}
\end{equation}
where the amplitude term gives no contribution to the closed Brillouin-zone integral. Therefore, whether a Lifshitz MCP carries a proper winding is determined by whether the phase factor $q_k$ can be smoothly defined at the gapless point.

\begin{figure}[t]
  \centering
  \begin{minipage}[b]{0.24\linewidth}
    \panelH{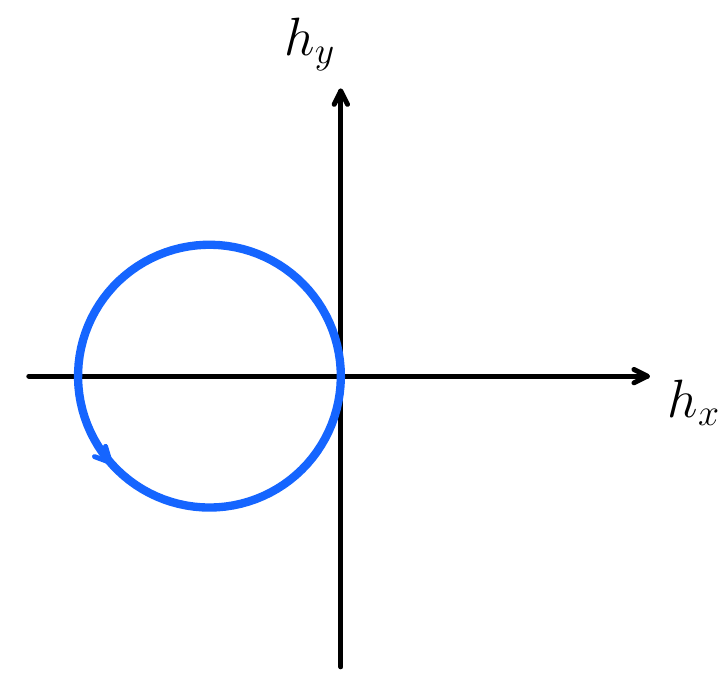}{4cm}{(a)}
  \end{minipage}
  \begin{minipage}[b]{0.23\linewidth}
    \panelH{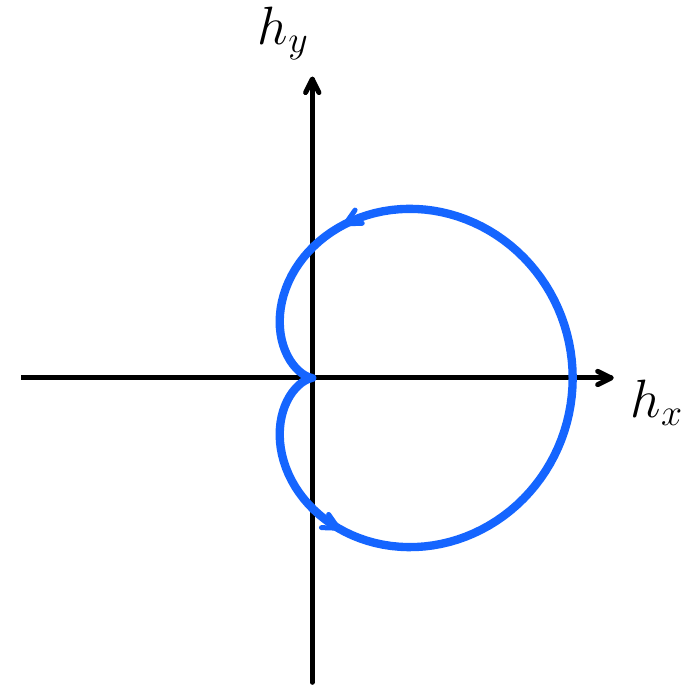}{4cm}{(b)}
  \end{minipage}
  \begin{minipage}[b]{0.27\linewidth}
    \panelH{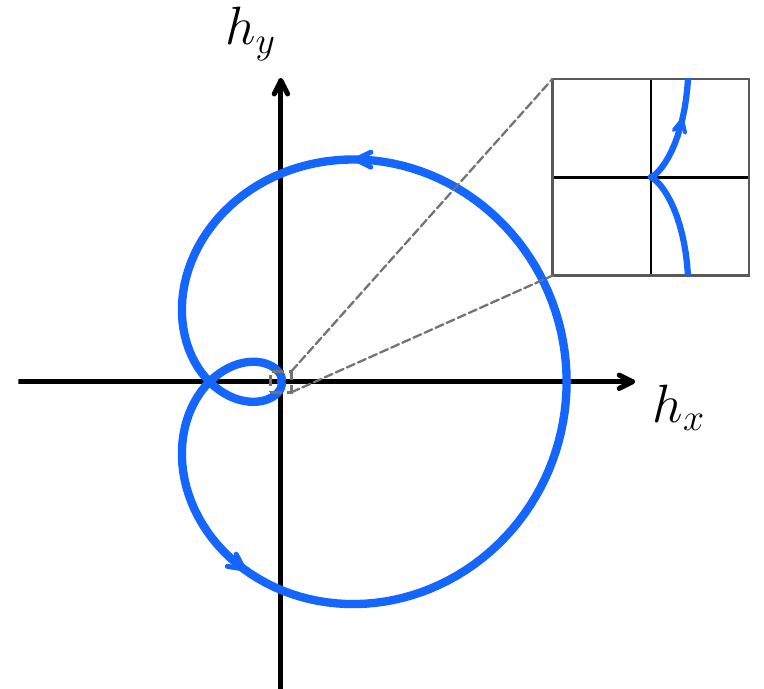}{4cm}{(c)}
  \end{minipage}
  \begin{minipage}[b]{0.22\linewidth}
    \panelH{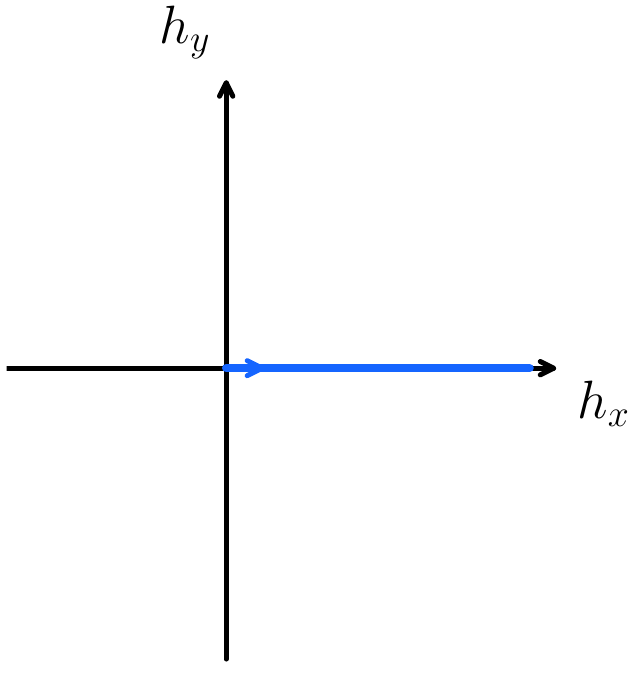}{4cm}{(d)}
  \end{minipage}
\caption{
Winding structure of the Lifshitz multicritical Bloch element in the $(h_x,h_y)$ plane.
(a) For the ordinary $z=1$ critical form $v_k=e^{ik}-1$, the continuous trajectory reaches the gapless point $v_k=0$, but the phase angle cannot be continued smoothly across this point; in the chosen branch, $\arg v_k$ jumps from $-\pi$ to $\pi$.
(b) For the even Lifshitz point $v_k=(e^{ik}-1)^2$, the trajectory also reaches $v_k=0$, but the phase angle is continuous across the Lifshitz zero. The flattened phase therefore remains smooth and carries winding $W=1$.
(c) For $v_k=(e^{ik}-1)^4$, the same even-$z$ mechanism gives a smooth phase evolution with winding $W=2$. The inset magnifies the neighborhood of $v_k=0$ and shows the local continuity of the raw $v_k$ trajectory near the Lifshitz zero.
(d) In the balanced even-$z$ case $v_k=e^{-ik}(1+e^{ik})^2=4\cos^2(k/2)$, the trajectory lies on the positive real axis, so the phase is trivial and $W=0$.
}
\label{fig:winding}
\end{figure}

For even $z=2m$, the $\alpha=0$ Lifshitz form gives
\begin{equation}
v_{2m}(k)
\sim
(e^{ik}-1)^{2m}=
\left(2i e^{ik/2}\sin\frac{k}{2}\right)^{2m}
\sim
e^{imk}
\left(\sin\frac{k}{2}\right)^{2m},
\end{equation}
up to an overall $k$-independent phase. Since $\left(\sin k/2\right)^{2m}$ is non-negative, it does not produce a sign jump across the gapless momentum, as illustrated by the comparison with the usual critical case in Fig.~\ref{fig:winding}(a,b). The flattened phase is therefore smooth and reduces to
\begin{equation}
q_k = e^{imk}.
\end{equation}
Substituting this into Eq.~\eqref{eq:winding_q}, we obtain
\begin{equation}
W=m=\frac{z}{2}.
\label{eq:even_winding_alpha0}
\end{equation}
Thus the even-$z$ Lifshitz MCP in the $\alpha=0$ family has a proper winding number, even though the unflattened spectrum remains gapless.

We next connect this winding to the PBC entanglement spectrum. For a chiral free-fermion ground state, the occupied-state correlation matrix is obtained from the flattened Hamiltonian. In momentum space, the occupied-state projector can be written as
\begin{equation}
P_-(k)=
\frac{1}{2}
\begin{pmatrix}
1 & -q_k^*\\
-q_k & 1
\end{pmatrix}.
\end{equation}
Fourier transforming $q_k$ gives the real-space flattened kernel
\begin{equation}
Q_{ij}=
\frac{1}{L}
\sum_k e^{ik(i-j)}q_k .
\end{equation}
After restricting the indices $i,j$ to the subsystem $A$, the correlation matrix takes the form
\begin{equation}
C_A=
\frac{1}{2}
\begin{pmatrix}
I_A & -Q_A^\dagger\\
-Q_A & I_A
\end{pmatrix},
\label{eq:CA_QA_even}
\end{equation}
where $Q_A$ is the subsystem-restricted matrix obtained from $Q_{ij}$.

For the even-$z$ Lifshitz MCP, $q_k\sim e^{imk}$, so
\begin{equation}
Q_{ij}\sim \delta_{j-i,m}.
\end{equation}
The flattened kernel is therefore an $m$-site shift. When this shift is restricted to a finite entanglement interval, $Q_A$ and $Q_A^\dagger$ each lose $m$ modes at the entanglement boundaries. These missing modes give eigenvalues of $C_A$ pinned at
\begin{equation}
\xi=\frac{1}{2}.
\end{equation}
Hence the number of midgap pairs is
\begin{equation}
N^{\rm ES}_{\rm pair}=m=\frac{z}{2}.
\label{eq:even_ES_pair_alpha0}
\end{equation}
This matches the pattern found in Eq.~\eqref{eq:pattern_even}, as well as the examples demonstrated for $z=2$ in the main text and for $z=4$ in Fig.~\ref{fig:numeric}(e). See Fig.~\ref{fig:winding}(d) for illustrate of $z=4$ case. Since these midgap modes are guaranteed by the winding number, as in the gapped SPT case, the midgap modes in even-$z$ Lifshitz multicritical points are topological rather than accidental.

The physical OBC spectrum is controlled by a different problem as presented in End Matter. Instead of the flattened phase $q_k$, one must solve the zero-mode equation of the unflattened Bloch element. For the $\alpha=0$ family, the boundary ansatz
\begin{equation}
\psi_j\sim \beta^j
\end{equation}
gives
\begin{equation}
v_z(\beta)\sim(\beta-1)^z=0 .
\end{equation}
Since the boundary ansatz is $\psi_j\sim \beta^j$, the root $\beta=1$ gives a spatial profile $\psi_j\sim 1$. Its amplitude remains unchanged as one moves away from the boundary into the bulk, and therefore the solution is not localized. Hence the Lifshitz root does not generate an OBC boundary zero mode. Thus
\begin{equation}
N^{\rm OBC}_{\rm pair }=0,
\end{equation}
This matches the pattern found in Eq.~\eqref{eq:pattern_even}, and consistent with the $z=2$ OBC spectrum in the main text and the $z=4$ OBC spectrum in Fig.~\ref{fig:numeric}(f).

Therefore, for even $z$, the Li--Haldane mismatch follows from the separation between the flattened and unflattened problems. The flattened phase $q_k$ is smooth and carries the winding $W=z/2$, producing $z/2$ PBC entanglement midgap pairs. By contrast, the unflattened Lifshitz root remains pinned at $\beta=1$ and does not generate localized physical OBC zero modes.

\subsection{General even-$z$ Lifshitz MCPs and hopping imbalance}

We now extend the above discussion from the $\alpha=0$ family to a general even-$z$ Lifshitz MCP between the $\alpha$ and $\alpha'$ critical lines. The multicritical Bloch element takes the form
\begin{equation}
v^{\rm mc}_{\alpha,\alpha'}(k)
\sim
e^{i\alpha k}(e^{ik}-1)^z,
\qquad
z=\alpha'-\alpha+1 .
\label{eq:vk_general}
\end{equation}
For even $z=2m$, we have
\begin{equation}
v^{\rm mc}_{\alpha,\alpha'}(k)
\sim
e^{i\alpha k}
\left(2i e^{ik/2}\sin\frac{k}{2}\right)^{2m}
\sim
e^{i(\alpha+m)k}
\left(\sin\frac{k}{2}\right)^{2m},
\end{equation}
up to an overall $k$-independent phase. Since the last factor is non-negative, it does not produce a sign jump across the gapless momentum. Therefore the flattened phase is smooth and reduces to
\begin{equation}
q_k
=
\frac{v^{\rm mc}_{\alpha,\alpha'}(k)}
{|v^{\rm mc}_{\alpha,\alpha'}(k)|}
=
e^{i(\alpha+m)k}.
\end{equation}
Substituting this into the winding formula gives
\begin{equation}
W
=
\frac{1}{2\pi i}
\int_{\rm BZ} dk\,
q_k^{-1}\partial_k q_k
=
\alpha+m
=
\frac{\alpha+\alpha'+1}{2}.
\label{eq:winding}
\end{equation}
This is the general winding number for even-$z$ Lifshitz MCPs mentioned in the main text, and it corresponds to the number of ES midgap pairs. 
\begin{equation}
    N_{\rm pair}^{\rm ES} = |W| = \frac{|\alpha+\alpha'+1|}{2}.
\end{equation}
We now compare this with the physical OBC zero-mode count, which is controlled by the unflattened root problem. For $\alpha\geq 0$, the multicritical Bloch element in Eq.~\eqref{eq:vk_general} contains $\alpha$ roots at $\beta=0$. These roots lie strictly inside the unit circle and therefore give $\alpha$ localized OBC zero-mode pairs. Hence
\begin{equation}
    N_{\rm pair}^{\rm OBC}
    =
    \alpha ,
    \qquad
    \alpha\geq 0 .
\end{equation}
Thus, for $\alpha\geq0$, the difference between the PBC entanglement midgap count and the physical OBC zero-mode count is
\begin{equation}
    N_{\rm pair}^{\rm ES}
    -
    N_{\rm pair}^{\rm OBC}
    =
    \frac{\alpha+\alpha'+1}{2}-\alpha
    =
    \frac{\alpha'-\alpha+1}{2}
    =
    \frac{z}{2}.
\end{equation}
This is the same additional contribution found in the $\alpha=0$ family. The additional $\alpha$ relative shift increases the ES midgap count and the OBC edge-mode count by the same number of pairs.

The mixed-sign case $\alpha<0$ and $\alpha'\geq0$ is more subtle because
$v^{\rm mc}_{\alpha,\alpha'}(\beta)$ is a Laurent polynomial. We therefore
follow the boundary-root counting rule described in the End Matter: for each
sublattice sector, we first remove the negative powers by multiplying by the
appropriate power of $\beta$, identify the roots strictly inside the unit
circle, and then subtract the boundary constraints generated by the removed
negative powers.

For the left-boundary $A$-sublattice zero-mode sector, the negative powers can be removed by multiplying by $\beta^{-\alpha}$. This gives
\begin{equation}
    \beta^{-\alpha}v^{\rm mc}_{\alpha,\alpha'}(\beta)
    \sim
    (\beta-1)^z .
\end{equation}
The resulting polynomial has no root strictly inside the unit circle; all roots remain at $\beta=1$. Including the $|\alpha|$ boundary constraints, the count is therefore
\begin{equation}
    N^{L}_{A}
    =
    \max\{0,\,0-|\alpha|\}
    =
    0 .
\end{equation}
For the left-boundary $B$-sublattice sector, the zero-mode equation is instead governed by the opposite chiral block $\bar v^{\rm mc}_{\alpha,\alpha'}(\beta)$. After removing the negative powers by multiplying by $\beta^{z-|\alpha|}$, we obtain
\begin{equation}
    \beta^{z-|\alpha|}\bar v^{\rm mc}_{\alpha,\alpha'}(\beta)
    \sim
    (\beta-1)^z .
\end{equation}
Again, all roots remain at $\beta=1$, so there is no root strictly inside the unit circle. The corresponding boundary count is
\begin{equation}
    N^{L}_{B}
    =
    \max\{0,\,0-(z-|\alpha|)\}
    =
    \max\{0,\,-\alpha'-1\}
    =
    0 .
\end{equation}
Thus the left boundary has no localized zero mode. The right boundary gives the same conclusion from the reversed boundary equations. Hence, for $\alpha<0$ and $\alpha'\geq0$,
\begin{equation}
    N_{\rm pair}^{\rm OBC}=0 .
\end{equation}
The difference between the PBC entanglement midgap count and the physical OBC zero-mode count is then
\begin{equation}
    N_{\rm pair}^{\rm ES}
    -
    N_{\rm pair}^{\rm OBC}
    =
    \frac{|\alpha+\alpha'+1|}{2}
    <
    \frac{\alpha'-\alpha+1}{2}
    =
    \frac{z}{2}.
\end{equation}
Thus, for fixed $z$, the mismatch is smaller than in the $\alpha=0$ case. This has a simple RG-limit interpretation: the maximum possible mismatch is set by half of the spatial width of the Lifshitz coupling, as discussed in SM Sec.~\ref{subsec:RGlimit}.

Finally, when both $\alpha$ and $\alpha'$ are negative, the problem can be mapped to the positive-range case by reversing the orientation, $\beta\to\beta^{-1}$. The same counting then applies, with the boundary chirality and edge location exchanged.

After establishing the Li--Haldane mismatch for general $\alpha,\alpha'$, we now return to the physical meaning of the winding number in Eq.~\eqref{eq:winding}. This result also has a simple real-space interpretation. In terms of $\beta=e^{ik}$, the multicritical Bloch element is
\begin{equation}
v^{\rm mc}_{\alpha,\alpha'}(\beta)
\sim
\beta^\alpha(\beta+1)^z, 
\end{equation}
when we place the gapless point at $k=\pi$. For $z=\alpha'-\alpha+1$, the hopping range extends from $\alpha$ to $\alpha+z=\alpha'+1$. The center of this finite-difference stencil is therefore
\begin{equation}
\frac{\alpha+(\alpha'+1)}{2}
=
W .
\end{equation}
Thus the winding number measures the imbalance of the hopping stencil. If the stencil is shifted away from the center, the flattened phase winds and the PBC entanglement spectrum contains topological midgap modes. If the stencil is centered, the net shift vanishes and the winding is zero.

As an illustrative example, we consider the balanced $z=2$ case. For the even Lifshitz MCP, the balanced condition corresponds to
\begin{equation}
    \alpha+\alpha'=-1,
\end{equation}
so the minimal example is $\alpha=-1$ and $\alpha'=0$. In this case, the two neighboring critical chains are generated by the symmetric factors
\begin{equation}
    v_{R}(k)\sim 1+e^{ik},
    \qquad
    v_{L}(k)\sim 1+e^{-ik},
\end{equation}
corresponding to the critical chains $H_0+H_1$ and $H_0+H_{-1}$, respectively. When the two critical roots collide, the multicritical Bloch element becomes
\begin{equation}
    v_{\rm bal}^{\rm mc}(k)
    \sim
    e^{-ik}(1+e^{ik})^2
    =
    2+e^{ik}+e^{-ik}
    =
    4\cos^2\frac{k}{2}.
\end{equation}
Thus the balanced multicritical Hamiltonian takes the real-space form
\begin{equation}
    H_{\rm bal}^{\rm mc}
    \sim
    \sum_j
    \left(
    a_j^\dagger b_{j-1}
    +
    2a_j^\dagger b_j
    +
    a_j^\dagger b_{j+1}
    +{\rm h.c.}
    \right).
\end{equation}
This is a symmetric second-difference Lifshitz operator, equivalent to the conventional Lifshitz Hamiltonian in Eq.~\eqref{eq:conven_vk} up to an overall sign and the choice of gapless momentum. This balanced construction is analogous to the $YY$--$ZZ$ Ising critical interface discussed in Ref.~\cite{prembabu2025noninvertibleinterfacessymmetryenrichedcritical}: both sides are topologically trivial in the edge-mode/ES-midgap sense, while differing only in their symmetry-enriched realization of the same critical theory.

From the flattened perspective, this balanced case has no net phase winding:
\begin{equation}
    q_k
    =
    \frac{v_{\rm bal}^{\rm mc}(k)}
    {|v_{\rm bal}^{\rm mc}(k)|}
    \sim 1,
    \qquad
    W=0 .
\end{equation}
This distinction is illustrated in Fig.~\ref{fig:winding}(d). The imbalanced even Lifshitz examples have nonzero winding and hence support PBC entanglement midgap modes, whereas the balanced even Lifshitz example has $W=0$ and is topologically trivial from the viewpoint of the flattened Hamiltonian.

\subsection{Odd-$z$ Lifshitz MCPs: relative shift and generalized Li--Haldane structure}

We next discuss the odd-$z$ Lifshitz MCPs in the $\alpha=0$ family. For $\alpha'=2m$, we have $z=2m+1$. The multicritical Bloch element is
\begin{equation}
    v_{2m+1}(k)
    \sim
    (e^{ik}-1)^{2m+1}.
\end{equation}
Using
\begin{equation}
    e^{ik}-1
    =
    2i e^{ik/2}\sin\frac{k}{2},
\end{equation}
we obtain
\begin{equation}
    v_{2m+1}(k)
    \sim
    e^{i(m+1/2)k}
    \left(\sin\frac{k}{2}\right)^{2m+1},
\end{equation}
up to an overall $k$-independent phase. Therefore the flattened phase contains the sign structure
\begin{equation}
    q_{2m+1}(k)
    =
    \frac{v_{2m+1}(k)}{|v_{2m+1}(k)|}
    \sim
    e^{i(m+1/2)k}
    {\rm sgn}\!\left(\sin\frac{k}{2}\right).
\end{equation}
Unlike the even-$z$ case, the phase angle of $q_k$ is discontinuous at the gapless momentum. Hence the ordinary winding number is not well defined for odd-$z$ Lifshitz MCPs.

Although the conventional winding is ill-defined, the entanglement calculation suggests that odd-$z$ Lifshitz MCPs share a similar long-range correlation structure with a CFT. To make this connection explicit, we compare the above flattened phase with that of the topological critical chain. The $\alpha$ critical chain is described by
\begin{equation}
    v_\alpha^{\rm c}(k)
    \sim
    e^{i\alpha k}(e^{ik}-1).
\end{equation}
After flattening, one obtains
\begin{equation}
    q_\alpha^{\rm c}(k)
    =
    \frac{v_\alpha^{\rm c}(k)}{|v_\alpha^{\rm c}(k)|}
    \sim
    e^{i(\alpha+1/2) k}
    {\rm sgn}\!\left(\sin\frac{k}{2}\right).
\end{equation}
Comparing this expression with the odd Lifshitz result, we see that the Lifshitz MCP between the $\alpha=0$ and $\alpha'=2m$ critical lines has the same flattened phase, and hence the same PBC correlation structure, as the $\alpha=m$ critical line.

We now use this identification to determine the entanglement midgap structure. For the $\alpha$ critical chain, the CFT structure makes the generalized Li--Haldane correspondence applicable to gapless SPT states, relating the PBC entanglement-spectrum midgap structure to the physical OBC boundary zero modes of the same critical chain \cite{guo2025generalizedlihaldanecorrespondencecritical}.

The OBC zero-mode number of the $\alpha$ critical chain can be counted directly from its unflattened root structure,
\begin{equation}
    v_\alpha^{\rm c}(\beta)
    \sim
    \beta^\alpha(\beta-1).
\end{equation}
The factor $\beta^\alpha$ gives $\alpha$ roots at $\beta=0$, which lie strictly inside the unit circle. Under the boundary ansatz $\psi_j\sim \beta^j$, these roots generate $\alpha$ localized zero modes at one boundary, together with the corresponding partners at the opposite boundary. Thus, in the pair-counting convention used for the entanglement spectrum,
\begin{equation}
    N_{\rm pair}^{\rm OBC}(\alpha)=\alpha .
\end{equation}
This count is topological because it is fixed by the number of roots inside the unit circle and cannot change under continuous symmetry-preserving deformations unless a root crosses the unit circle. The generalized Li--Haldane correspondence then gives
\begin{equation}
    N_{\rm pair}^{\rm ES}(\alpha)
    =
    N_{\rm pair}^{\rm OBC}(\alpha)
    =
    \alpha .
\end{equation}
Since the odd Lifshitz MCP between the $\alpha=0$ and $\alpha'=2m$ critical lines has the same PBC correlation structure as the $\alpha=m$ critical line, it contains
\begin{equation}
    N_{\rm pair}^{\rm ES}
    =
    m
    =
    \frac{z-1}{2}
\end{equation}
entanglement midgap pairs. This is consistent with the pattern found in Eq.~\eqref{eq:pattern_odd} and explains the $z=3$ PBC correlation spectrum in Fig.~\ref{fig:numeric}(c), where one midgap pair appears.

We finally compare this with the physical OBC spectrum of the odd Lifshitz MCP itself. The physical OBC zero-mode problem is controlled by the unflattened Lifshitz Bloch element
\begin{equation}
    v_{2m+1}(\beta)
    \sim
    (\beta-1)^{2m+1}.
\end{equation}
There is no $\beta^m$ factor and hence no root at the origin. The only root is $\beta=1$, with multiplicity $2m+1$. Since the boundary ansatz is $\psi_j\sim \beta^j$, the root $\beta=1$ gives a spatial profile $\psi_j\sim 1$. Its amplitude remains unchanged as one moves away from the boundary into the bulk, and therefore the solution is not localized. Hence
\begin{equation}
    N_{\rm pair}^{\rm OBC}=0,
\end{equation}
consistent with the pattern found in Eq.~\eqref{eq:pattern_odd} and with the $z=3$ OBC spectrum shown in Fig.~\ref{fig:numeric}(d).

This gives the odd-$z$ version of the Li--Haldane mismatch. The PBC entanglement spectrum is governed by the flattened phase and is identical to that of an $\alpha=m$ gapless SPT critical chain, while the physical OBC spectrum is governed by the unflattened Lifshitz root structure and contains no localized zero mode.

The same reasoning extends directly to a general odd-$z$ Lifshitz MCP. Restoring the overall factor $e^{i\alpha k}$, the multicritical Bloch element is
\begin{equation}
    v^{\rm mc}_{\alpha,\alpha'}(k)
    \sim
    e^{i\alpha k}(e^{ik}-1)^{2m+1},
    \qquad
    z=2m+1,
    \qquad
    \alpha'=\alpha+2m .
\end{equation}
After flattening, this gives
\begin{equation}
    q^{\rm mc}_{\alpha,\alpha'}(k)
    \sim
    e^{i(\alpha+m+1/2)k}
    {\rm sgn}\!\left(\sin\frac{k}{2}\right).
\end{equation}
This is the same flattened phase as the symmetry-enriched critical chain with effective label
\begin{equation}
    \alpha_{\rm eff}
    =
    \alpha+m
    =
    \frac{\alpha+\alpha'}{2},
\end{equation}
which is an integer because $\alpha'-\alpha=2m$. Therefore the PBC entanglement spectrum of the general odd Lifshitz MCP is the same as that of the $\alpha_{\rm eff}$ critical chain. Using the generalized Li--Haldane correspondence and the root-counting result of the $\alpha$ critical chain, the number of ES midgap pairs is
\begin{equation}
    N_{\rm pair}^{\rm ES}
    =
    N_{\rm pair}^{\rm ES}(\alpha_{\rm eff})
    =
    \min\left\{
    \left|\frac{\alpha+\alpha'}{2}\right|,
    \left|\frac{\alpha+\alpha'}{2}+1\right|
    \right\}.
\end{equation}

The physical OBC zero-mode count, however, is again determined by the unflattened Lifshitz root problem
\begin{equation}
    v^{\rm mc}_{\alpha,\alpha'}(\beta)
    \sim
    \beta^\alpha(\beta-1)^{2m+1}.
\end{equation}
For $\alpha\geq0$, the $\beta^\alpha$ factor gives $\alpha$ localized OBC zero-mode pairs, while the Lifshitz factor only gives roots at $\beta=1$ and does not contribute localized boundary modes. Thus
\begin{equation}
    N_{\rm pair}^{\rm OBC}=\alpha,
    \qquad
    \alpha\geq0.
\end{equation}
The difference is then
\begin{equation}
    N_{\rm pair}^{\rm ES}
    -
    N_{\rm pair}^{\rm OBC}
    =
    m
    =
    \frac{z-1}{2}.
\end{equation}
Compared with the $\alpha=0$ case, the additional overall shift by $\alpha$ increases the ES midgap count and the OBC edge-mode count by the same number of pairs.

For the mixed-sign case $\alpha<0$ and $\alpha'\geq0$, the same sector-dependent boundary-root counting used above gives no localized OBC zero modes,
\begin{equation}
    N_{\rm pair}^{\rm OBC}=0.
\end{equation}
The difference is then,
\begin{equation}
    N_{\rm pair}^{\rm ES}
    -
    N_{\rm pair}^{\rm OBC}
    =
    \min\left\{
    \left|\frac{\alpha+\alpha'}{2}\right|,
    \left|\frac{\alpha+\alpha'}{2}+1\right|
    \right\}
    <
    \frac{\alpha'-\alpha}{2}
    =
    \frac{z-1}{2}.
\end{equation}
Again, the mismatch is smaller than in the $\alpha=0$ case, with the same physical interpretation as in the even-$z$ Lifshitz discussion.

Finally, when both $\alpha$ and $\alpha'$ are negative, the problem can be mapped to the positive-range case by reversing the orientation, $\beta\to\beta^{-1}$. The same counting then applies, with the boundary chirality and edge location exchanged.

\subsection{RG-limit coupling-range criterion for the OBC spectrum}
\label{subsec:RGlimit}

\begin{figure}[t]
\centering
\includegraphics[width=1.0\textwidth]{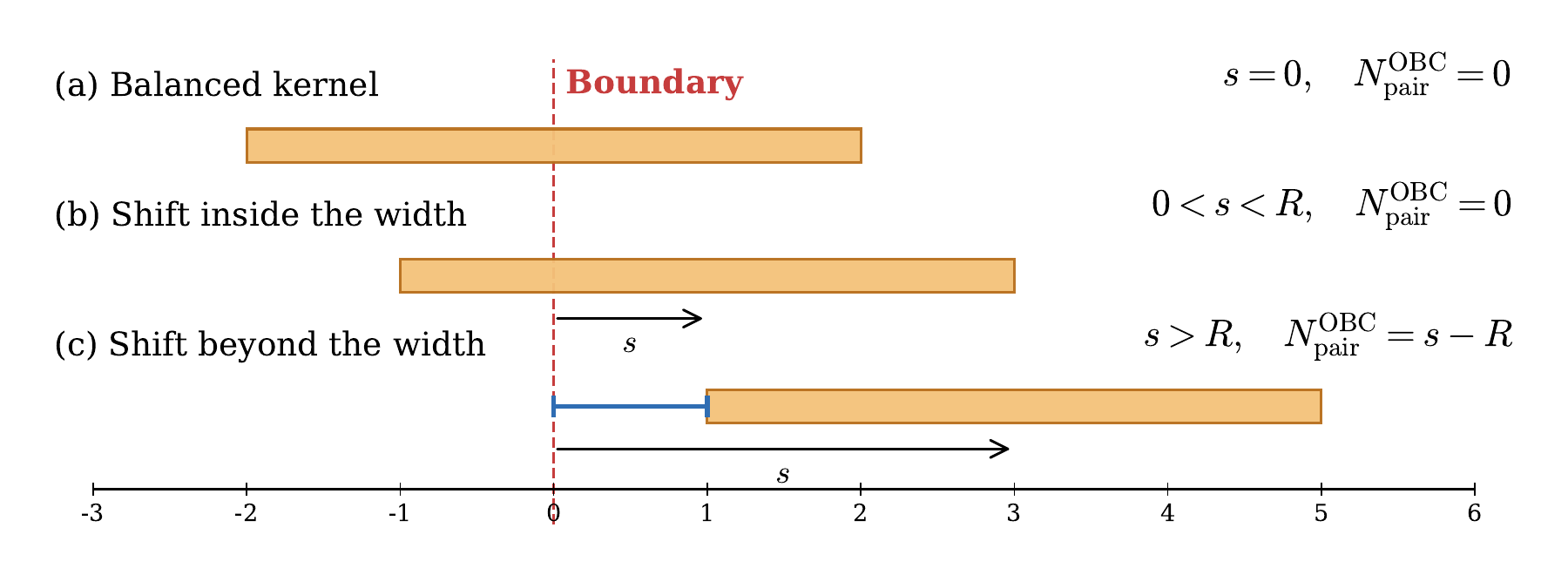}
\caption{
RG-limit coupling-range criterion for the OBC edge-mode count.
The orange bar denotes the real-space coupling range after a relative displacement $s$, and the dashed red line marks the physical boundary.
For clarity, the schematic is drawn for an even-width coupling, corresponding to the even-$z$ Lifshitz case; the same coupling-range logic also applies to odd-$z$ Lifshitz MCPs.
The shift $s$ is the quantity detected by the flattened correlation kernel and therefore corresponds to the PBC entanglement midgap-pair count.
(a) For a balanced kernel, $s=0$, the coupling range is centered around the boundary and no boundary channel is decoupled.
(b) When the shift remains inside the half-width, $0<s<R$, the entanglement spectrum detects the shifted structure, but the physical OBC coupling still reaches the boundary, so no localized OBC edge mode is produced.
(c) When the shift exceeds the half-width, $s>R$, part of the shifted coupling range moves past the boundary, giving $N_{\rm pair}^{\rm OBC}=s-R$ localized edge-mode pairs.
Thus the finite width of the Lifshitz coupling explains why the PBC entanglement spectrum can detect a shifted topological structure, while the physical OBC spectrum changes only when the shifted coupling range moves beyond the boundary.
}
\label{fig:LH_sch}
\end{figure}

The OBC counting above can be summarized by a simple coupling-range criterion in the RG-limit picture. Consider a centered real-space coupling stencil whose nonzero matrix elements extend from $-R$ to $R$,
\begin{equation}
\mathcal I_\Delta=[-R,R],
\end{equation}
where $R$ is the half-width of the coupling range. After applying a relative shift by $s$, the coupling range becomes
\begin{equation}
\mathcal I_{T^s\Delta}=[s-R,s+R].
\end{equation}
The entanglement cut is sensitive to this relative shift in the flattened kernel. By contrast, a physical OBC edge mode requires an actual decoupled boundary degree of freedom in the unflattened real-space Hamiltonian. In the RG-limit cartoon, this means that the shifted coupling range must move past the physical boundary. As illustrated in Fig.~\ref{fig:LH_sch}, for a positive shift the number of decoupled boundary channels is therefore
\begin{equation}
N_{\rm pair}^{\rm OBC}
=
\max\{0,s-R\}.
\label{eq:support_width_obc}
\end{equation}
The opposite orientation is obtained by reflecting the coupling range.

This criterion immediately explains why the usual Li--Haldane correspondence works for ordinary SPT chains but can fail for Lifshitz MCPs. For a gapped SPT fixed point, the RG-limit coupling has zero width, $R=0$. A unit shift $s=1$ already moves the whole coupling away from the boundary and creates one dangling boundary degree of freedom. Thus the shift seen by the entanglement cut is also visible as a physical OBC edge mode.

For a Lifshitz MCP, however, the coupling has finite spatial width. The simplest even example has $R=1$ and a natural shift $s=1$, corresponding to Fig.~\ref{fig:LH_sch}(b). In this case the shifted coupling range still reaches the boundary, so Eq.~\eqref{eq:support_width_obc} gives no decoupled boundary channel. Thus the flattened kernel can carry one unit of winding, producing an ES midgap pair, while the physical OBC Hamiltonian still has no localized boundary zero mode. If the shift is further increased, as in Fig.~\ref{fig:LH_sch}(c), the part $s-R$ moves beyond the boundary and gives the physical OBC edge-mode count.

The same criterion also explains the general $\alpha,\alpha'$ counting. For the even-$z$ Lifshitz form
\begin{equation}
v^{\rm mc}_{\alpha,\alpha'}(\beta)
\sim
\beta^\alpha(\beta-1)^z,
\end{equation}
the coupling range extends from $\alpha$ to $\alpha+z$. Hence
\begin{equation}
R=\frac{z}{2},
\qquad
s=\alpha+\frac{z}{2}.
\end{equation}
Equation~\eqref{eq:support_width_obc} then gives
\begin{equation}
N_{\rm pair}^{\rm OBC}
=
\max\{0,\alpha\},
\end{equation}
which is precisely the root-counting result derived above for the positive- and mixed-sign cases. The flattened winding, on the other hand, counts the shift $s$. Therefore, for $\alpha\geq0$, the extra $\alpha$ shift increases both the ES midgap count and the OBC edge-mode count by the same amount, while the remaining mismatch is fixed by the half-width $R=z/2$. For mixed-sign cases, the shifted coupling range only partially moves away from the boundary, so the mismatch is smaller than this maximal value, as schematically represented by Fig.~\ref{fig:LH_sch}(b).

For odd-$z$ Lifshitz MCPs, the flattened phase does not define a proper winding number. Nevertheless, the same OBC coupling-range criterion applies to the unflattened Lifshitz coupling. The PBC entanglement spectrum can inherit midgap modes from the corresponding symmetry-enriched critical kernel, as discussed above, but the physical OBC spectrum is still controlled by whether the finite-width Lifshitz coupling leaves an actual dangling boundary channel. When the shifted coupling range still reaches the boundary, the physical OBC zero mode is absent.

Thus the Li--Haldane breakdown at topologically enforced Lifshitz MCPs can be viewed as a finite-width boundary effect. The flattened kernel can retain winding or relative-shift topology and therefore produce PBC entanglement midgap modes. However, the physical OBC edge modes are controlled by the full real-space range of the unflattened Lifshitz stencil. A relative shift that is visible in the entanglement spectrum does not necessarily decouple a boundary degree of freedom unless it exceeds the half-width of the coupling range.

\end{document}